\documentclass[journal]{IEEEtran}
\usepackage{latexsym,amsmath,amstext,amsfonts,amssymb,graphicx,epsfig}
\usepackage{algorithm, algorithmic, multirow, booktabs}
\usepackage{times}
\usepackage{flushend}
\usepackage{subfigure}
\usepackage{epstopdf}
\usepackage{latexsym}
\usepackage{color}
\usepackage{xcolor}%
\usepackage{float}%
\usepackage{colortbl,booktabs}%

\ifCLASSINFOpdf
\else
\fi
\begin{document}
%
\title{Study of Robust Two-Stage Reduced-Dimension Sparsity-Aware STAP with Coprime Arrays}
%
%
%

\author{Xiaoye~Wang,
        Zhaocheng~Yang,~\IEEEmembership{Member,~IEEE,} \\
        ~Jianjun~Huang,~\IEEEmembership{Member~IEEE,}
        ~and~Rodrigo C.~de~Lamare,~\IEEEmembership{Senior Member~IEEE}%

\thanks{Xiaoye Wang and Zhaocheng Yang are with the Guangdong Key Laboratory of Intelligent Information Processing,
College of Electronics and Information Engineering, Shenzhen University, Shenzhen, 518060, P.R. China (e-mail: wangxiaoye2013@126.com, yangzhaocheng@szu.edu.cn).
Jianjun Huang is with the College of Electronics and Information Engineering, Shenzhen University, Shenzhen, 518060, P.R. China (e-mail: huangjin@szu.edu.cn).
R. C. de Lamare is with the Department of Electronics, University of York, YO10
5DD, York, UK (e-mail: delamare@cetuc.puc-rio.br). }
\thanks{This work was supported in part by National Natural Science Foundation of China (61771317), Guangdong Basic and Applied Basic Research Foundation (2019A1515011517), and Science and Technology Project of Shenzhen (JCYJ20170302150111535), the Natural Science Foundation of SZU (827-000236).}}

%
%

\markboth{Submitted to IEEE Transactions on Signal Processing ~2019}%
{Shell \MakeLowercase{\textit{et al.}}: Bare Demo of IEEEtran.cls for IEEE Journals}
%



\maketitle

\begin{abstract}
Space-time adaptive processing (STAP) algorithms with coprime arrays can provide good clutter suppression potential with low cost in airborne radar systems as compared with their uniform linear arrays counterparts. However, the performance of these algorithms is limited by the training samples support in practical applications. To address this issue, a robust two-stage reduced-dimension (RD) sparsity-aware STAP algorithm is proposed in this work. In the first stage, an RD virtual snapshot is constructed using all spatial channels but only $m$ adjacent Doppler channels around the target Doppler frequency to reduce the slow-time dimension of the signal. In the second stage, an RD sparse measurement modeling is formulated based on the constructed RD virtual snapshot, where the sparsity of clutter and the prior knowledge of the clutter ridge are exploited to formulate an RD overcomplete dictionary. Moreover, an orthogonal matching pursuit (OMP)-like method is proposed to recover the clutter subspace. In order to set the stopping parameter of the OMP-like method, a robust clutter rank estimation approach is developed. Compared with recently developed sparsity-aware STAP algorithms, the size of the proposed sparse representation dictionary is much smaller, resulting in low complexity. Simulation results show that the proposed algorithm is robust to prior knowledge errors and can provide good clutter suppression performance in low sample support.
\end{abstract}


%
\IEEEpeerreviewmaketitle

\section{Introduction}

 Space-time adaptive processing (STAP) has received substantial attention since its inception due to its potential in offering improved performance for clutter suppression and target detection \cite{WardReport1994}. It is known that
the full-dimension (FD) STAP algorithm requires at least two times the degrees of freedom (DoFs) of independent and identically distributed (IID) samples to achieve a signal-to-interference-plus-noise ratio (SINR) loss within 3dB compared to the optimum performance, which is usually impractical for heterogenous environments, especially with large arrays. Furthermore, high computational complexity and storage space are needed to compute the FD STAP filter \cite{MelvinHetero2000}. Therefore, STAP algorithms with attractive performance at low sample support and low computational complexity are of great importance in practical applications.

To deal with such issues, numerous STAP algorithms have been
proposed in the last decades
\cite{YLWVari2000,HWOnAdp1994,YWangRobust2003,HaiEig1996,GoldsteinRR1997,
GoldsteinMultsg1998,SarkarDeter2001,Knowledge2006,GuerciKA2006}.
Reduced-dimension (RD) STAP algorithms, such as the factored
algorithm (FA) or extended FA (EFA) \cite{WardReport1994}, multiple
Doppler channels joint processing scheme (mDT) \cite{YLWVari2000},
joint-domain localized algorithm (JDL) \cite{HWOnAdp1994}, and
space-time multiple beam (STMB) algorithm \cite{YWangRobust2003},
are proposed by employing a low dimension for reducing the
computational complexity and sample support requirements. These
algorithms have limited steady-state performance due to reduced
system DoFs. In this context, reduced-rank algorithms
\cite{Vantrees1,locsme,elnashar,manikas,cgbf,okspme,r19,scharf,bar-ness,pados99,
reed98,hua,goldstein,santos,qian,delamarespl07,delamaretsp,xutsa,xu&liu,
kwak,delamareccm,delamareelb,wcccm,delamarecl,delamaresp,delamaretvt,delamaretvt10,delamaretvt2011ST,
delamare_ccmmswf,jidf_echo,jidf,sjidf,barc,lei09,delamare10,fa10,ccmavf,lei10,jio_ccm,
ccmavf,stap_jio,zhaocheng,zhaocheng2,arh_eusipco,arh_taes,rdrab,dcg,dce,dta_ls,
song,wljio,barc,saalt,jiodoa,rrdoa,damdc,kaesprit,mmimo,wence,spa,mbdf,rrmber,bfidd,did,mbthp,rmbthp,wlbd,joint,bbprec,baplnc,wcccm,locsme,rabtaes,locsmeiet,okspme,lrcc},
such as the principle components (PC) \cite{HaiEig1996},
cross-spectral metric (CSM) \cite{GoldsteinRR1997}, multistage
Wiener filter (MSWF) \cite{GoldsteinMultsg1998}, and the joint
interpolation, decimation and filtering (JIDF) \cite{jidf,sjidf} can
provide high steady-state performance by using two times the clutter
rank of IID samples. Direct-data domain (D3) STAP algorithms are
developed to bypass sample support problems by only using the
received data of the cell under test (CUT) \cite{SarkarDeter2001}.
However, the advantage of these algorithms comes at the expense of
reduced system DoFs. Recently, knowledge-aided STAP algorithms have
shown improved performance with a small number of training samples
by exploiting the prior knowledge of environments or radar systems
\cite{Knowledge2006,GuerciKA2006}. However, these algorithms suffer
from performance degradation in presence of prior knowledge errors.

Owing to the successful application of compressive sensing in the parameter estimation, recent attention has been focusing on sparsity-aware STAP (termed SA-STAP) techniques by exploiting the sparsity of the clutter \cite{YangClut2013,SenLow2015,SunDSR2011,FengSBL2015,WuSBL2016,DuanSpars2017,WangClut2017,ZYEnhanced2017, KnowSR2016,ZCYangSP2019,HanSpect2017,SparsS2017,GuoSparRd2017,KDoffGrid2018}. In the SA-STAP algorithms, the clutter components are first represented by an overcomplete dictionary multiplying a sparse vector (i.e., the clutter spectrum) in the angle-Doppler plane, and can be efficiently estimated via sparse recovery techniques.
The SA-STAP algorithms show advantages of fast convergence speed and high resolution of parameter estimation. However, they are costly due to spectrum grid search, especially when the dimension of the sparse vector is high. To overcome this problem, several low complexity SA-STAP algorithms have been proposed \cite{KnowSR2016,ZYEnhanced2017,HanSpect2017,GuoSparRd2017,ZCYangSP2019}.
In presence of imperfections, several studies have also been devoted to robust beamforming and robust SA-STAP algorithms \cite{Khabbazibasmenj2012,RCDL2016,SparsS2017,KnowSR2016,ZYEnhanced2017}.

It is worth noting that the STAP algorithms aforementioned generally assume that both the space and slow-time samples are gathered by Nyquist sampling. However, because of various considerations such as weight, power, configuration and electronic counterpart, Nyquist sampling is hardly employed \cite{SparASTAP1998, AThinSTAP1999,Vaidyanathan2011}. Hence, sparse sampling has been brought to airborne radar applications, where the effective DoFs are reduced and only limited performance can be achieved. More recently, motivated by attractive advantages including large aperture, low mutual coupling, and increased DoFs in the virtual domain provided by the coprime sampling \cite{Vaidyanathan2011,CopSTAP2015,YujieGu2018,ChunLLCRB2017,MianzhiWCMCRB2017}, several STAP algorithms have been developed for clutter suppression using coprime sampling configurations \cite{WangSTAP2018,wxySparse2018}.
Two STAP algorithms based on virtual construction and the spatial-temporal smoothing technique for coprime arrays (CPAs) have been proposed in \cite{WangSTAP2018}, where a much reduced number of array elements and pulses is used while the performance is close to that of a standard array with Nyquist sampling. However, the advantage comes at the cost of an increase in the number of training samples. In addition, the developed algorithms in \cite{WangSTAP2018} do not make full use of the total DoFs offered by the derived virtual snapshot. In order to overcome these problems, we proposed an FD SA-STAP (FD-SA-STAP) algorithm for clutter suppression using all DoFs of the virtual snapshot \cite{wxySparse2018}, which can achieve much better clutter suppression performance in very low sample support than the counterpart in \cite{WangSTAP2018}. However, the high computational complexity of the FD-SA-STAP algorithm is still a challenge for real-time applications.

In this paper, a robust two-stage RD SA-STAP considering inaccurate prior knowledge (RTSKA-RD-SA-STAP) is presented. In the proposed RTSKA-RD-SA-STAP algorithm, preprocessing by the discrete Fourier transform (DFT) with regard to the slow-time for each sensor is adopted. An RD clutter covariance matrix is estimated by using all spatial channels but only $m$ adjacent Doppler channels around the target Doppler frequency, which is similar to the mDT. Since the obtained RD covariance matrix is no longer a Toeplitz matrix, the conventional virtual transformation operation developed in \cite{WangSTAP2018,wxySparse2018} can not be directly applicable. To this end, the diagonal block matrices of the resultant RD covariance matrix, which have Toeplitz structure, instead of the whole RD covariance matrix \cite{wxySparse2018}, are used to derive an RD virtual snapshot and the relationship between the RD virtual snapshot and the FD virtual snapshot is derived. Then, the sparse measurement model of the constructed RD virtual snapshot is formulated. Here, another RD overcomplete dictionary is proposed by using the prior knowledge of the clutter ridge, and an orthogonal matching pursuit (OMP)-like method is developed to recover the clutter subspace. To set the stopping parameter of the OMP-like method, a robust clutter rank estimation approach is developed for the CPA by considering errors in prior knowledge. Hence, an eigenanalysis-based method can be applied to design the STAP filter. Additionally, the convergence, the implementations and computational complexity of RTSKA-RD-SA-STAP are also analyzed. Simulations are provided to demonstrate the theoretical derivations and advantages of the proposed RTSKA-RD-SA-STAP algorithm. The main contributions of this paper are summarized as follows:

1) We propose a robust two-stage RD SA-STAP algorithm denoted as RTSKA-RD-SA-STAP for improving the performance of clutter suppression in airborne radars with CPAs under limited training sample support.

2) We develop a robust clutter rank estimation approach based on inaccurate prior knowledge which enables the setting of parameters in the RTSKA-RD-SA-STAP algorithm. It is also shown that the proposed clutter rank estimation approach is applicable to both side-looking and non-side-looking airborne radars.

3) We analyze the convergence of the virtual construction, where the relationship between the FD virtual construction and RD virtual construction is established, and discuss the feasibility of the proposed RTSKA-RD-SA-STAP from practical implementation and computational complexity perspectives.

The rest of this paper is organized as follows. The signal model and problem formulation are introduced in Section \ref{Problem}. The proposed RTSKA-RD-SA-STAP algorithm is detailed in Section \ref{Proposed}. The convergence analysis, the implementations and computational complexity of RTSKA-RD-SA-STAP are discussed in \ref{analysis}. In Section \ref{Simulation}, numerical examples are conducted to demonstrate the performance of RTSKA-RD-SA-STAP, and finally, conclusion is given in Section \ref{Conclusion}.
%

\section{ Signal Model and Problem Formulation } \label{Problem}
\subsection{Signal Model} \label{SigModel}
\begin{figure}[!htbp]
\centering
\includegraphics[width=2.1in]{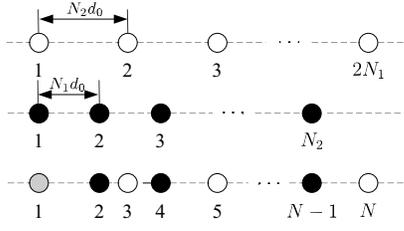}
\caption{The configuration of CPA.}
\label{coprimeArr}
\end{figure}

Consider an airborne radar system employing an $N$-sensor array and an $M$-pulse train in a coherent processing interval (CPI). The height and velocity of the airborne platform are $h_p$ and $v_p$, respectively. The sensors of the array are employed as a CPA, as shown in Fig. \ref{coprimeArr}. Here, $N_1$ and $N_2$ are coprime pair integers ($N_1<N_2$), $d_0$ is the half wavelength, and $N = 2N_1+N_2-1$. The radar transmits the pulse train at a fixed pulse repetition interval (PRI) $T_r$. Considering a target located at a given range bin, the received return is usually modeled as
\begin{equation}\label{recSig}
\begin{split}
    {\bf x}=  \alpha_{t} {\bf v}(\varpi_t,\vartheta_t) + {\bf x}_{u},
\end{split}
\end{equation}
where $\alpha_t$ is the unknown complex amplitude of the target, ${\bf v}(\varpi_t,\vartheta_t)$ is a space-time steering vector corresponding to the normalized
Doppler frequency $\varpi_t$ and spatial frequency $\vartheta_t$ of the target, and ${\bf x}_{u}$ is the interference component. For given frequencies $\varpi$ and $\vartheta$, the space-time steering vector is given by $ {\bf v}(\varpi,\vartheta) = {\bf b} (\varpi) \otimes {\bf a} (\vartheta) $, where
\begin{equation}\label{temST}
\begin{split}
  {\bf b}(\varpi)  = [1, e^{j2 \pi \varpi},\cdots, e^{j 2 \pi (M-1) \varpi} ]^T,
\end{split}
\end{equation}
and
\begin{equation}\label{spST}
\begin{split}
 {\bf a}(\vartheta)  = [1, e^{j 2 \pi { d_2 } \vartheta},\cdots, e^{j2 \pi d_N  \vartheta}]^T,
\end{split}
\end{equation}
are the temporal steering vector and spatial steering vector, respectively, $d_i, i=2,\cdots,N$ denotes the position of the $i\textrm{th}$ sensor with respect to the first array sensor, $(\cdot)^T$ denotes the transpose operator, and $\otimes$ represents the Kronecker product. The interference component ${\bf x}_u$ is assumed to consist of clutter and noise and is usually modeled as
\begin{equation}\label{intf}
\begin{split}
  {\bf x}_u
  = {\bf x}_{c} + {\bf n}= \sum_{i=1}^{N_c} \alpha_{c,i} {\bf v}(\varpi_{c,i},\vartheta_{c,i}) + {\bf n}= {\bf V}{\boldsymbol \alpha}_c + {\bf n},
\end{split}
\end{equation}
where ${\bf x}_c$ is the clutter component, ${\bf n}$ is the thermal noise of the radar receiver, $N_c$ is the number of IID clutter patches in a given range bin, $\alpha_{c,i}$ is the complex amplitude of $i\textrm{th}$ clutter patch, ${\boldsymbol \alpha}_c  = [\alpha_{c,1}, \cdots, \alpha_{c,N_c}]^T$, and ${\bf V} = [{\bf v}(\varpi_{c,1},\vartheta_{c,1}),\cdots,{\bf v}({\varpi}_{c,N_c},\vartheta_{c,N_c})]$.

The STAP filter weight vector that maximizes the output SINR is given by \cite{WardReport1994}
\begin{equation}\label{weight}
\begin{split}
  {\bf w} = \frac{{\bf R}^{-1} {\bf v}(\varpi_t, \vartheta_t)}{{\bf v}^H(\varpi_t, \vartheta_t){\bf R}^{-1} {\bf v}(\varpi_t, \vartheta_t)},
\end{split}
\end{equation}
where $(\cdot)^{-1}$ denotes the matrix inversion operator and ${\bf R}$ is the interference covariance, computed by
\begin{equation}\label{Cov}
\begin{split}
  {\bf R}
   =  E \{ {\bf x}_u {\bf x}^H_u \}
  = {\bf V} \textrm{diag}({\bf p}) {\bf V}^H + \sigma^2_n {\bf I}.
\end{split}
\end{equation}
Here, ${\bf p} = E \{ {\boldsymbol \alpha}_c \circ{\boldsymbol \alpha}^H_c \} = [|\alpha|^2_{c,1},\cdots,|\alpha|^2_{c,N_c}]^T$ with $\circ $ being the Schur-Hadamard product operator and $|\cdot|$ being the absolute operator, $E \{\cdot \}$ denotes the mathematical expectation, $\textrm{diag}(\cdot)$ denotes a diagonal matrix composed of the bracketed elements, $(\cdot)^H$ denotes the Hermitian transpose, $\sigma^2_n$ is the power of thermal noise, and ${\bf I}$ denotes an identity matrix with appropriate dimension. In practice, the covariance matrix ${\bf R}$ is estimated from a set of training samples ${\bf x}_i,i=1,\cdots,L$ by
\begin{equation}\label{CovHat}
\begin{split}
  {\bf \hat R} =  \frac{1}{L} \sum^L_{i=1}  {\bf x}_i {\bf x}^H_i,
\end{split}
\end{equation}
where $L$ is the total number of training samples.

\subsection{FD-SA-STAP \cite{wxySparse2018} }

\begin{figure}[!htbp]
\centering
\includegraphics[width=90mm]{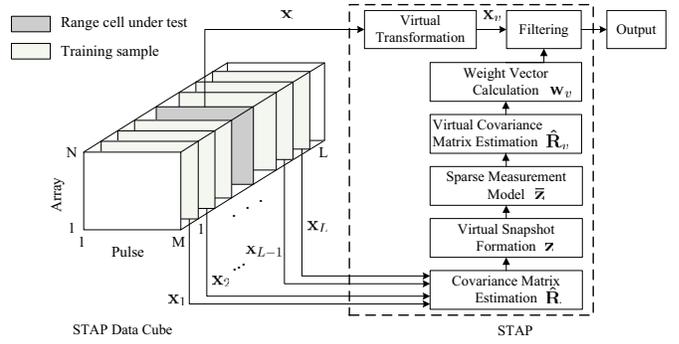}
\caption{The procedure of the FD-SA-STAP algorithm.}
\label{FullDimSASTAP}
\end{figure}

Taking advantage of the properties of increased DoFs and large aperture offered by the CPA as well as the fast convergence of the sparse recovery algorithms, an FD-SA-STAP algorithm was recently proposed to suppress the clutter in airborne radar with CPAs \cite{wxySparse2018}. Its corresponding procedure is shown in Fig. \ref{FullDimSASTAP}.

As seen in Fig. \ref{FullDimSASTAP}, the filter output in the virtual domain is given by
 \begin{equation}\label{equivaSnapshotVc}
 {\bf y}_v = {\bf w}^H_v {\bf x}_v,
\end{equation}
 where ${\bf x}_v$ is the virtual signal at the CUT and ${\bf w}_v$ is the filter weight vector in the virtual domain derived by three main steps: 1) constructing the virtual snapshot ${\bf z}$ by applying a transformation including vectorization, removing redundancy, and rearrangement on the covariance matrix estimate; 2) forming the sparse measurement model ${\bf \bar{z} }$ and estimating the virtual covariance matrix ${\bf \hat R}_v$ based on the sparse measurement model via sparse recovery techniques; 3) computing the STAP filter weight vector ${\bf w}_v $ by the virtual covariance matrix estimate. Here, the resultant virtual snapshot ${\bf z}$, its sparse measurement model ${\bf \bar{z} }$, and the STAP filter weight vector ${\bf w}_v $ are expressed as
\begin{equation}\label{equivaSnapshotVc}
 {\bf z} = {\bf V}_v {\bf p} + {\sigma}^2_n {\bf e}_0,
\end{equation}
\begin{equation}\label{diffCoarraySR}
    {\bf \bar{z} } = {\bf \Phi} {\bf \bar{p} } + {\sigma}^2_n {\bf e}_0,
\end{equation}
\begin{equation}\label{filter}
    {\bf w}_v = \frac{ {\bf \hat R}^{-1}_{v} {\bf v}_v({\varpi}_t,{\vartheta}_t) } {{\bf v}^H_v({\varpi}_t,{\vartheta}_t) {\bf \hat R}^{-1}_v {\bf v}_v({\varpi}_t,{\vartheta}_t) } ,
\end{equation}
where ${\bf V}_v$ denotes the virtual steering matrix of the clutter, ${\bf \Phi}$ is the $N_vM_v \times N_dN_s$ space-time steering dictionary matrix with $N_v$ being the number of virtual sensors, $M_v$ being the number of virtual pulses, and $N_d$ and $N_s$ being respectively the number of Doppler bins and angle bins in angle-Doppler plane, ${\bf \bar{p}}$ is the sparse vector of the clutter, ${\bf e}_0$ is a column vector of all zeros except for a one at the central position, and ${\bf v}_v({\varpi}_t,{\vartheta}_t)$ denotes the virtual space-time steering vector of the target which can be obtained by
\begin{equation}\label{filter}
   {\bf v}_v({\varpi}_t,{\vartheta}_t) = {\bf H} ({\bf v}^* ({\varpi}_t,{\vartheta}_t) \otimes {\bf v} ({\varpi}_t,{\vartheta}_t)).
\end{equation}
Here, ${\bf H}$ is the virtual transformation matrix and $(\cdot)^*$ denotes the complex conjugation operator. Assumed the clutter spectrum estimate ${\bf \hat {\bar p}}$, the virtual covariance matrix ${\bf \hat R}_v$ is given by
\begin{equation}\label{filter}
   {\bf \hat R}_v = {\bf \Phi} \textrm{diag}({\bf \hat {\bar p}}) {\bf \Phi}^H + {\hat \sigma}^2_n {\bf I}.
\end{equation}
The virtual signal ${\bf x}_v$, corresponding to the received signal at the CUT in \eqref{recSig}, can be obtained as
\begin{equation}\label{filter}
  {\bf x}_v = {\bf H} ({\bf x}^* \otimes {\bf x}) = {\bf z}_t + {\bf z},
\end{equation}
where
\begin{equation}\label{filter}
  {\bf z}_t = |\alpha_t|^2 {\bf v}_v({\varpi}_t,{\vartheta}_t).
\end{equation}
Then, the corresponding output SINR is defined by
\begin{equation}\label{SINR}
\begin{split}
\textrm{SINR} = \frac{|\alpha_t|^2 |{\bf w}^H_{v} {\bf v}_{v}(\varpi_t, \vartheta_t)|^2 }{{\bf w}^H_{v} {\bf R}_{v} {\bf w}_{v}}.
\end{split}
\end{equation}
For more details of the FD-SA-STAP algorithm, interested readers are referred to \cite{wxySparse2018}.

The determination of ${\bf \hat R}_v$ requires high computational complexity since both the dimension of the problem and the number of atoms in the dictionary matrix ${\bf \Phi}$ are large. Moreover, the derived STAP filter weight vector involves a high complexity operation, i.e., the inversion of the covariance matrix ${\bf \hat  R}_v$, whose complexity is $ \mathcal{O}(N_v M_v)^3$. To this end, we propose a low complexity RTSKA-RD-SA-STAP algorithm in the sequel.

\section{Proposed RTSKA-RD-SA-STAP Algorithm} \label{Proposed}
In this section, we detail the proposed RTSKA-RD-SA-STAP algorithm in three parts: the first RD stage constructs the RD virtual snapshot; the second RD stage performs an RD sparse measurement modeling considering inaccurate prior knowledge; and the clutter subspace estimation and STAP filter design.

\subsection{The First RD Stage: RD Virtual Snapshot Construction}
First, we apply the DFT with regard to the slow-time for each sensor to reduce the dimension of the radar received signal. Using the RD transformation matrix ${\bf U}$, we express the RD signal as
\begin{equation}\label{rdRecSig}
\begin{split}
    {\bf x}_r
    =  {\bf U}^H  {\bf x} = \alpha_{t} {\bf U}^H {\bf v}(\varpi_t,\vartheta_t) + {\bf U}^H  {\bf x}_{u}.
\end{split}
\end{equation}
For clarity, the term ${\bf U}^H  {\bf x}_{u}$ in \eqref{rdRecSig} can be rewritten as
\begin{equation}\label{rdIntf}
\begin{split}
  {\bf x}_{r,u} = {\bf U}^H{\bf x}_{u} = {\bf U}^H {\bf x}_{c} + {\bf U}^H {\bf n}\textcolor{red}{.}
\end{split}
\end{equation}
It is assumed in this paper that the RD transformation is only applied to the signal in the Doppler domain. Hence, the matrix ${\bf U}$ is given by
\begin{equation}\label{rdMat}
\begin{split}
  {\bf U} = {\bf U}_t \otimes {\bf I},
\end{split}
\end{equation}
where ${\bf U}_t$ denotes the $M \times m$ DFT matrix in the Doppler domain with $m$ being the number of Doppler bins selected, given by
\begin{equation}\label{rdMatTim}
\begin{split}
  {\bf U}_t = [{\bf u} ({\varpi}_1),{\bf u} (\varpi_2),\cdots,{\bf u} (\varpi_m) ].
\end{split}
\end{equation}
Here, ${\bf u} ({\varpi}_i), i= 1,2,\cdots, m $ is an $M \times 1$ vector from the DFT matrix at the frequency ${\varpi}_i, i= 1,2,\cdots, m $, given by
\begin{equation}\label{rdMatTim}
\begin{split}
  {\bf u} ({\varpi}_i) = [1,e^{j 2 \pi {\varpi}_i},\cdots,e^{j 2 \pi  (M-1) {\varpi}_i}]^T,
\end{split}
\end{equation}
and ${\varpi}_i = {\varpi}_t + (i-1 - \frac{m-1}{2}) \frac{1}{M} $. From \eqref{rdIntf}, the covariance matrix of the RD signal takes the form
\begin{equation}\label{rdCov}
\begin{split}
{\bf R}_r
  & =\left[
  \tiny
     \begin{matrix}
     E\{ {\bf s}(\varpi_1){\bf s}^H(\varpi_1) \} & E\{ {\bf s}(\varpi_1){\bf s}^H(\varpi_2) \} & \cdots & E\{ {\bf s}(\varpi_1){\bf s}^H(\varpi_m) \}\\
     E\{ {\bf s}(\varpi_2){\bf s}^H(\varpi_1) \} & E\{ {\bf s}(\varpi_2){\bf s}^H (\varpi_2)\} & \cdots & E\{ {\bf s}(\varpi_2){\bf s}^H(\varpi_m) \}\\
     \cdots &\cdots & \cdots & \cdots \\
      E\{ {\bf s}(\varpi_m){\bf s}^H(\varpi_1) \} & E\{ {\bf s}(\varpi_m){\bf s}^H (\varpi_2)\} & \cdots & E\{ {\bf s}(\varpi_m){\bf s}^H(\varpi_m) \}\\
     \end{matrix}
     \right],
\end{split}
\end{equation}
where ${\bf s}(\varpi_i),i=1,\cdots,m$ are the DFT coefficients associated with the received signal $\bf x_u$ at the Doppler frequency $\varpi_i$.

Note that the covariance matrix ${\bf R}_r $ does not have a Toeplitz structure, and hence, the existing virtual snapshot construction process is not directly applicable to the covariance matrix ${\bf R}_r $. Fortunately, the block Toeplitz structure is maintained for the covariance matrix ${\bf R}_r $. In other words, for a single Doppler frequency $\varpi$, the term $E\{ {\bf s}(\varpi){\bf s}^H(\varpi) \}$ is a Toeplitz matrix which contains the whole spatial information of interference component ${\bf x}_u$ at the Doppler bin $\varpi$. More exactly, according to \eqref{rdCov}, the diagonal block matrices of ${\bf R}_r $ is stacked as follows
\begin{equation}\label{rdStaCov}
\begin{split}
  {\bf \Pi} =\left[
     \begin{matrix}
      E\{ {\bf s}(\varpi_1){\bf s}^H(\varpi_1) \}\\
      E\{ {\bf s}(\varpi_2){\bf s}^H (\varpi_2)\} \\
     \cdots\\
      E\{ {\bf s}(\varpi_m){\bf s}^H(\varpi_m) \}\\
     \end{matrix}
     \right],
\end{split}
\end{equation}
Because each block matrix in \eqref{rdStaCov} is a Toeplitz matrix, the traditional virtual snapshot construction can be straightforwardly applied. The $i$th virtual subsnapshot associated with the $i$th Doppler bin can be expressed as
\begin{equation}\label{rdVirSp_k}
\begin{split}
  {\bf z}_{r,i} = {\bf P} \textrm{vec}(E\{ {\bf s}(\varpi_i){\bf s}^H (\varpi_i)\}),
\end{split}
\end{equation}
where $\textrm{vec}(\cdot)$ represents the vectorized form of a matrix, ${\bf P}$ is the $N_v \times N^2$ virtual transformation matrix, and its detailed information is shown in Appendix \ref{apdx1}. After the virtual transformation for all diagonal block matrices of ${\bf R}_r $, the virtual snapshot can be finally written as
\begin{equation}\label{rdVirSp}
\begin{split}
  {\bf z}_r =\left[
     \begin{matrix}
      {\bf z}_{r,1}  \\
      {\bf z}_{r,2}  \\
     \cdots\\
      {\bf z}_{r,m}  \\
     \end{matrix}
     \right]
    = \left[
     \begin{matrix}
      {\bf P} \textrm{vec}(E\{ {\bf s}(\varpi_1){\bf s}^H (\varpi_1)\}) \\
      {\bf P} \textrm{vec}(E\{ {\bf s}(\varpi_2){\bf s}^H (\varpi_2)\})\\
     \cdots\\
      {\bf P} \textrm{vec}(E\{ {\bf s}(\varpi_m){\bf s}^H(\varpi_m) \}) \\
     \end{matrix}
     \right] .
\end{split}
\end{equation}
We note that $\textrm{vec}(E\{ {\bf s}(\varpi_i){\bf s}^H(\varpi_i) \})$ can be represented as
\begin{equation}\label{rdVirSp_k1}
\begin{split}
  \textrm{vec}(E\{ {\bf s}(\varpi_i){\bf s}^H(\varpi_i) \}) = \textrm{vec}({\bf u}^H(\varpi_i) {\bf R} {\bf u}(\varpi_i ) ).
\end{split}
\end{equation}
Using the property of the Kronecker product $\textrm{vec}({\bf a} {\bf A} {\bf b}^T) = ({\bf b} \otimes {\bf a} ) \textrm{vec}({\bf A})$, we rewrite \eqref{rdVirSp_k1} as
\begin{equation}\label{rdVirSp_k2}
\begin{split}
  \textrm{vec}(E\{ {\bf s}(\varpi_i){\bf s}^H (\varpi_i)\}) = ({\bf u}^T(\varpi_i) \otimes {\bf u}^H(\varpi_i )) {\bf r},
\end{split}
\end{equation}
where ${\bf r} = \textrm{vec} ({\bf R})$. By applying this relation to all virtual snapshots in ${\bf z}_r $ of \eqref{rdVirSp}, we have
\begin{equation}\label{rdVirSp1}
\begin{split}
  {\bf z}_r = \left[
     \begin{matrix}
      {\bf P} ({\bf u}^T(\varpi_1) \otimes {\bf u}^H(\varpi_1 )) {\bf r} \\
      {\bf P} ({\bf u}^T(\varpi_2) \otimes {\bf u}^H(\varpi_2 )) {\bf r} \\
     \cdots\\
      {\bf P} ({\bf u}^T(\varpi_m) \otimes {\bf u}^H(\varpi_m )) {\bf r} \\
     \end{matrix}
     \right].
\end{split}
\end{equation}
As shown in\cite{ChunLLCRB2017,MianzhiWCMCRB2017}, and \cite{WangSTAP2018}, the vector $ {\bf r}$ in \eqref{rdVirSp1} can be expressed in terms of ${\bf z}$ in \eqref{equivaSnapshotVc} as
\begin{equation}\label{virSPCovRel}
\begin{split}
  {\bf z} = {\bf F } {\bf r},
\end{split}
\end{equation}
where ${\bf F}$ is an $N_v M_v \times N^2M^2 $ matrix which establishes the relationship between the vectorized form ${\bf r}$ of the covariance matrix ${\bf R}$ and the FD virtual snapshot vector ${\bf z}$ (see Appendix \ref{apdx2} for definition of $\bf F$).
Since the matrix ${\bf F}$ has a pseudo-inverse, one can substitute \eqref{virSPCovRel} into \eqref{rdVirSp1} and get
\begin{equation}\label{rdVirSp2}
\begin{split}
  {\bf z}_r = {\bf G} {\bf z}.
\end{split}
\end{equation}
where
\begin{equation}\label{rdVirSp2}
\begin{split}
  {\bf G} = \left[
     \begin{matrix}
      {\bf P} ({\bf u}^T(\varpi_1) \otimes {\bf u}^H(\varpi_1 )) \\
      {\bf P} ({\bf u}^T(\varpi_2) \otimes {\bf u}^H(\varpi_2 )) \\
     \cdots\\
      {\bf P} ({\bf u}^T(\varpi_m) \otimes {\bf u}^H(\varpi_m )) \\
     \end{matrix}
     \right] {\bf F }^\dag,
\end{split}
\end{equation}
and $(\cdot)^{\dag}$ denotes the pseudo inversion operator. This means that the virtual snapshot ${\bf z}_r$ for the RD signal can be directly determined from the virtual snapshot ${\bf z}$ for the FD signal.

It should be noted that if $m$ is set equal to the number of pulses $M$, the resultant method is a particular case of the one where the FD virtual transformation is performed on the Doppler-element domain data. The performance of the FD virtual transformed STAP derived from the Doppler-element domain data is similar to that of FD-SA-STAP. In order to reduce the training sample support and the computational complexity, $m$ with a value being much smaller than $M$ is more appropriate according to the partial adaptive processing perspectives shown in \cite{WardReport1994} and \cite{HWOnAdp1994}. More precisely, to ensure an efficient computational complexity and achievable performance in limited training sample support, it is shown that three or four Doppler bins are selected and a performance trade-off can be achieved from the post-Doppler processing STAP works in \cite{WardReport1994} and \cite{HWOnAdp1994}. Thus, following the same ideas, we set $m$ to 3. Moreover, in the simulations, we shall see that the proposed RTSKA-RD-SA-STAP algorithm with more than three Doppler bins (i.e. $m > 3$) has very little performance improvement. Therefore, the proposed RTSKA-RD-SA-STAP algorithm with $m=3$ may be a more attractive trade-off and be applicable for practical radar applications.

According to \eqref{diffCoarraySR}, the sparse model of ${\bf z}_r$ in \eqref{rdVirSp1} can be expressed as
\begin{eqnarray}\label{srRdVirSp4}
\begin{split}
  {\bf \bar{z}}_{r} = {\bf \Phi }_r {\bf \bar{p}} + {\sigma}_n^2 {\bf G} {\bf e}_0,
\end{split}
\end{eqnarray}
where ${\bf \Phi }_r = {\bf G} {\bf \Phi}$ stands for the $mN_v \times N_dN_s$ RD dictionary matrix.

Note that although the length of atoms in the RD dictionary matrix ${\bf \Phi }_r$ is reduced from $N_vM_v$ to $m N_v$, the cost is relatively high since the number of atoms in the dictionary matrix ${\bf \Phi }_r$ is $N_dN_s$. In order to deal with this problem, by exploiting the fact that the clutter is located around the clutter ridge, a second RD stage is derived to formulate an RD sparse measurement model by using prior knowledge of the clutter ridge, even in presence of prior knowledge errors.

\subsection{The Second RD Stage: RD Sparse Measurement Modeling Considering Inaccurate Prior Knowledge}
Now, let us proceed to the second RD stage by formulating the RD sparse measurement model for the clutter. For a given range bin, it is known that the location of a clutter patch in airborne radars is described by the spatial frequency ${\vartheta}$ and Doppler frequency ${\varpi}$, given by
\begin{equation}\label{spaFreq}
\begin{split}
  {\vartheta} = \frac{d_0}{\lambda} \cos  {\varphi} \sin {\theta},
\end{split}
\end{equation}
\begin{equation}\label{timFreq}
\begin{split}
{\varpi} = \frac{2 v_p} {\lambda} \cos {\varphi} \sin ({\theta} + {\psi}).
\end{split}
\end{equation}
where ${\varphi}$, ${\theta}$, and ${\psi}$ are the elevation angle, azimuth angle, and crab angle, respectively. For notation simplicity, the subscript 'c' is removed. Note that these frequencies in \eqref{spaFreq} and \eqref{timFreq} assume an exactly known radar system parameters, such as elevation angle ${\varphi}$, azimuth angle ${\theta}$, and crab angle ${\psi}$ (${\psi}= 0$ for side-looking arrays).
However, this assumption could be violated due to imperfect measurements and array errors in practice. The authors in \cite{ZYEnhanced2017} discussed the impact of practical imperfections on these radar system parameters, reported that the practical imperfections show little effects on elevation angles $\varphi$ while much more effects on platform velocity $v_p$ and crab angle ${\psi}$, and modeled the actual Doppler frequency measured by uniform linear array (ULA) to have an uncertainty $\Delta {\varpi}$ which is bounded as $| \Delta {\varpi} | \leq \Delta {\varpi}_m$, where $\Delta {\varpi}$ is the uncertainty of the actual Doppler frequency. Inspired by the Doppler frequency uncertainty in ULA radars, we exploit this uncertainty in a similar manner for the CPA radars, and denote the actual Doppler frequency by
\begin{eqnarray}\label{difTimFreq}
\begin{split}
{\varpi}_{\textrm{actual}} = {\varpi}+ \Delta {\varpi},
\end{split}
\end{eqnarray}
where ${\varpi}$ is the assumed Doppler frequency. Assuming the error's range values of the platform velocity $\Delta v_{pm}$ and of the crab angle $\Delta {\psi}_m$, the $\Delta {\varpi}$ satisfies the following bound \cite{ZYEnhanced2017}:
\begin{eqnarray}\label{difTimFreq2}
\begin{split}
  | \Delta {\varpi} |
  & \leq \frac{2 }{\lambda} | \cos {\varphi} | \\
  & \times \sqrt{ [\frac{v'_p }{2} (\Delta \psi_{m})^2 + v'_p -1]^2 +  \Delta v^2_{pm}  - (v'_p -1)^2 }   \\
  & =  \Delta {\varpi}_m,
\end{split}
\end{eqnarray}
where $v'_p$ is the measured platform velocity which may contain measured errors. As a result, $ \Delta {\varpi}_m $ gives the bounds of the uncertainties caused by imperfect knowledge of radar parameters. It should be noted that the right hand side of \eqref{difTimFreq2} represents deterministic error bounds of Doppler frequencies that are related to ${\varphi} $, $v'_p$, $\Delta \psi_{m}$, and $\Delta v_{pm}$, which can be obtained from the inertial navigation unit (INU), Global Positioning Satellite (GPS) data, and previous known experience \cite{Knowledge2006,ZYEnhanced2017}. This implies that the actual Doppler frequency of a clutter should lie within a region around the true Doppler frequency, i.e.
\begin{eqnarray}\label{TimFreq3}
\begin{split}
  {\varpi}_{ \textrm{actual} } \in [{\varpi}-\Delta {\varpi}_m , {\varpi} + \Delta {\varpi}_m  ].
\end{split}
\end{eqnarray}
We propose to approximately estimate the actual Doppler frequency of a clutter patch by partitioning the period $[  {\varpi}-\Delta {\varpi}_m, {\varpi} + {\Delta} {\varpi}_m  ]$ into a group of $M_e$ grid points $\{ {\varpi}'_{i} \}^{M_e}_{i=1}$ uniformly. That is, the actual Doppler frequency should lie in the set
\begin{eqnarray}\label{timFreqk}
\begin{split}
 \Omega = \{ {\varpi}_{min}, {\varpi}_{min} + \Delta_d, \cdots, {\varpi}_{min} + (M_e-1) \Delta_d \},
\end{split}
\end{eqnarray}
where $ {\varpi}_{min} = {\varpi} - \Delta {\varpi}_m$ and $\Delta_d = \frac{2 \Delta {\varpi}_m}{M_e-1}$.

According to the above discussions, for a given range bin, we divide the range of the azimuth angle uniformly into $ N'_c$ grid points $\{ \theta_i \}^{ N'_c}_{i=1}$. Then, the $i$th corresponding spatial frequency can be expressed as \eqref{spaFreq} with $\theta$ replaced by $\theta_i$ and the $i$th Doppler frequency related to \eqref{timFreq} is replaced by the set $\Omega_i$, which is obtained via \eqref{timFreqk} with ${\varpi}$ replaced by ${\varpi}_{i} = \frac{2 v_p}{\lambda} \cos {\varphi} \sin ({\theta}_i + {\psi})$. Stacking the Doppler frequencies corresponding to all azimuth angles, one gets
\begin{eqnarray}\label{timFreqkall}
\begin{split}
 \bar{\Omega} = \{  \Omega_1,  \Omega_2, \cdots,  \Omega_{ N'_c}\} = \{ \bar{{\varpi}}_1, \bar{{\varpi}}_2,\cdots, \bar{{\varpi}}_{{M}_d }\},
\end{split}
\end{eqnarray}
where $\Omega_i, i=1,\cdots, N'_c$ denotes the $i$th possible Doppler frequency set associated with the $i$th azimuth angle, and $\bar{{\varpi}}_i,i=1,\cdots, M_d$ represents the $i$th Doppler frequency in the formed set $\bar{\Omega}$ with $ M_d = M_e  N'_c$ being the number of the resultant discretized Doppler frequency bins.

\begin{figure}[!htbp]
\centering
\includegraphics[width=90mm]{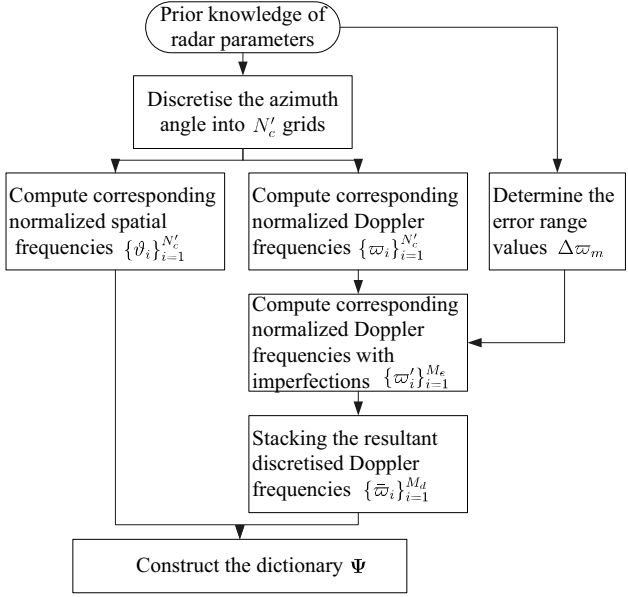}
\caption{The procedure of the overcomplete dictionary construction.}
\label{dictionary}
\end{figure}

For notational simplicity, assuming $ \{ \bar{{\varpi}}_i \}^{M_d}_{i=1}$ is the resultant discretized Doppler frequency bins. Consequently, we can construct an $mN_v \times M_d$ two-stage RD dictionary $ \bf \Psi $ according to these discretised spatial frequencies and Doppler frequencies. Different from the method suggested in \cite{KDoffGrid2018} based on exact prior knowledge of radar parameters, the proposed method for the dictionary construction considers the imperfections of knowledge of radar system parameters. The detailed procedure of constructing dictionary $ \bf \Psi $ is summarized in Fig. \ref{dictionary}.

Finally, according to \eqref{srRdVirSp4}, the two-stage RD sparse measurement model can be reformulated as
 \begin{eqnarray}\label{srRdVirSp5}
\begin{split}
 {\bf \tilde{z}}_{r} = {\bf \Psi} {\bf \tilde{p}} + {\sigma}_n^2 {\bf G} {\bf e}_0.
\end{split}
\end{eqnarray}
 In practice, since the virtual snapshot is derived from the covariance matrix estimate $\bf \hat R$, the ${\bf \tilde{z}}_{r}$ in \eqref{srRdVirSp5} is thus expressed as
\begin{eqnarray}\label{srRdVirSp6}
\begin{split}
  {\bf \hat{\tilde{z}}}_{r} = {\bf \Psi} {\bf \tilde{p}} + {\sigma}_n^2 {\bf G} {\bf e}_0 + {\boldsymbol \epsilon},
\end{split}
\end{eqnarray}
where ${\boldsymbol \epsilon}$ stands for the estimation error of the virtual snapshot.

Note that mismatches may exist between the real clutter and the dictionary matrix ${\bf \Phi}$ in \eqref{diffCoarraySR} or ${\bf \Psi}$ \eqref{srRdVirSp6} as the clutter is continuously distributed. Such errors or mismatches are referred to as the off-grid problem by many references such as \cite{YangClut2013} and \cite{ZYang2013}. Compared with the off-grid problem in sparsity-based direction-of-arrival estimation (DOA) approaches, the SA-STAP algorithm is less sensitive to the off-grid problem. More exactly, in sparsity-based DOA estimation approaches, the problem of DOA estimation is formulated as a sparse recovery problem where the support of the sparse signal to be recovered is just the DOAs of interest. This implies that the sparsity-based DOA estimation approaches sample the range of angles of interest onto fixed sampling grids, which serve as the set of all candidates of DOA estimates, and assume that all true unknown DOAs are exactly on the selected grid \cite{ZYang2013}. However, for the proposed RTSKA-RD-SA-STAP algorithm,  the main goal is to suppress the clutter and not to estimate the positions of the clutter. More precisely, for the proposed RTSKA-RD-SA-STAP algorithm, the clutter subspace rather than the exact positions of the clutter component is the primary focus. Since the atoms in the overcomplete dictionary have some relevance \cite{XYang2019}, a suitable set of the space-time steering vectors from the dictionary can be selected to accurately estimate the clutter subspace \cite{YangClut2013}. Therefore, we are able to reconstruct the clutter subspace by selecting some atoms from the overcomplete dictionary.

We should also note that the overcomplete dictionary formulation is critical to the performance of SA-STAP. Intuitively, the larger the number of dictionary atoms, the better the performance of the recovered solutions but the higher the computational complexity required by sparse reconstruction. However, the performance improvement is very small while the computational complexity dramatically increases when the number of dictionary atoms is larger than some value \cite{YangClut2013,XYang2019}. In the simulations, we will detail this problem.

\subsection{ The Clutter Subspace Estimation and STAP Filter Design }

According to the sparsity of the clutter, the spectrum of the clutter can be estimated by solving the following minimization problem
\begin{eqnarray}\label{srRdVirSp7}
\begin{split}
  \mathop{{\min}} \limits_{ {\bf \tilde{p}} } \| {\bf \tilde{p}} \|_1  \quad s.t. \ \| {\bf \hat{\tilde{z}}}_{r} - {\bf \Psi} {\bf \tilde{p}}- {\sigma}_n^2 {\bf G} {\bf e}_0 \|_2 \leq\zeta_1,
\end{split}
\end{eqnarray}
where $\| \cdot \|_{i} (i=1,2)$ is the $l_i$ norm and $\zeta_1$ characterizes virtual snapshot estimation error. Additionally, it is assumed there is no target signal in the training samples. If the target signal occurs in the training samples, the existing training data selection methods can be applied to select the training samples before employing the proposed algorithm \cite{KGerlach2002}.

The problem in \eqref{srRdVirSp7} can be solved by a number of well-known sparse recovery algorithms, such as OMP \cite{JOMP2007}, the least absolute shrinkage and selection operator (LASSO) method \cite{RLASSO1996}, focal underdetermined system solver (FOCUSS) \cite{IFOCUSS2007} and sparse Bayesian learning (SBL) \cite{MSBL2001}. For the purpose of reducing the computational complexity, an OMP-like method is developed to solve this optimization problem. Moreover, in order to further reduce the complexity, an eigenanalysis-based method is adopted to calculate the STAP filter weight vector by employing the clutter subspace estimate rather than the clutter covariance matrix estimate.

First, the clutter subspace can be estimated by the proposed OMP-like method, which is summarized in Table \ref{cluSpaEst}. Then the estimated clutter subspace can be represented by
\begin{eqnarray}\label{cluSpa5}
\begin{split}
  {\bf \hat  v }_{rv,k} = \frac{[{\bf \Psi}]_{:,i_k} - \sum^{k-1}_{j=1} {\bf \hat v }^H_{rv,j} [{\bf \Psi}]_{:,i_k} {\bf \hat v }_{rv,j} }{ \|[{\bf \Psi}]_{:,i_k} - \sum^{k-1}_{j=1} {\bf \hat v }^H_{rv,j} [{\bf \Psi}]_{:,i_k} {\bf \hat v }_{rv,j} \|_2 },
\end{split}
\end{eqnarray}
where $i_k$ is the index set at the $k$th iteration, ${\bf \hat v }_{rv,j},j=1,\cdots,k-1$ denotes the selected steering vectors after $k-1$ iterations, and $[{\bf \Psi}]_{:,k}$ denotes the $k$th column of ${\bf \Psi}$. If given an appropriate number of iterations $K$, we can obtain the clutter subspace via the set of steering vectors ${\bf \hat V}_{rv,K}$. Finally, the STAP filter weight vector based on the eigenanalysis-based method can be determined by
\begin{eqnarray}\label{filter}
\begin{split}
    {\bf w}_{rv} = ({\bf I} - {\bf \hat V}_{rv} {\bf \hat V}^H_{rv})  {\bf v}_{rv}({\varpi}_t,{\vartheta}_t),
\end{split}
\end{eqnarray}
where ${\bf v}_{rv}({\varpi}_t,{\vartheta}_t) = {\bf G}{\bf v}_{v}({\varpi}_t,{\vartheta}_t)$ is the RD virtual space-time steering vector of the target, and ${\bf \hat V}_{rv}={\bf \hat V}_{rv,K}$ is the clutter subspace estimate. Finally, the whole procedure of the proposed RTSKA-RD-SA-STAP algorithm is summarized in Table \ref{CAUP RV InAME-KA}.
\begin{table}[!htbp]
  \centering
  \caption{Proposed method for clutter subspace estimation}\label{cluSpaEst}
  \begin{tabular}{l}
  \hline
  1 \textbf{Initialization:} \\
   \quad ${\bf \gamma }_0 = {\bf z}_r $, $K = N_{cx,2}$, ${\lambda}_0 = \emptyset$, ${\bf \hat V }_{rv,0} = \emptyset$, ${\bf \Psi}_0 = {\bf \Psi}$. \\
  2 \textbf{Find:}  \\
   \quad  $i_1 = \textrm{argmax}  |{\bf \Psi}^H {\bf \gamma }_0 |$, ${\bf \hat v}_{rv,1} = \frac{ [{\bf \Psi}]_{:,i_1} }{ \|[{\bf \Psi}]_{:,i_1} \|_2 } $. \\
  3 \textbf{Update:} \\
   \quad ${\lambda}_1 = \{ i_1 \}$, ${\bf \Psi}_1 = {\bf \Psi}_0 - [{\bf \Psi}]_{:,i_1} $, \\
    \quad ${\bf \gamma }_1 = {\bf \gamma }_0  - {\bf \hat v}^H_{rv,1} {\bf \gamma }_0 {\bf \hat v}_{rv,1}  $, ${\bf \hat V}_{rv,1} = {\bf \hat V }_{rv,0} \bigcup {\bf \hat v}_{rv,1} $. \\
  4 \textbf{For $k=2:K$ } \\
  \quad $i_k = \textrm{argmax}  |{\bf \Psi}^H {\bf \gamma }_{k-1} | $,\\
  \quad ${\bf \hat v}_{rv,k} = \frac{[{\bf \Psi}]_{:,i_k} - \sum^{k-1}_{j=1} {\bf \hat v}^H_{rv,j} [{\bf \Psi}]_{:,i_k} {\bf \hat v}_{rv,j} }{ \|[{\bf \Psi}]_{:,i_k} - \sum^{k-1}_{j=1} {\bf \hat v}^H_{rv,j} [{\bf \Psi}]_{:,i_k} {\bf \hat v}_{rv,j} \|_2 } $, \\
  \quad ${\lambda}_k = {\lambda}_{k-1} \bigcup i_{k} $, ${\bf \Psi}_{k} = {\bf \Psi}_{k-1} - [{\bf \Psi}]_{:,i_k} $, \\
  \quad ${\bf \gamma }_k = {\bf \gamma }_{k-1}  - {\bf \hat v}^H_{rv,k} {\bf \gamma }_{k-1} {\bf \hat v}_{rv,k}  $, ${\bf \hat V}_{rv,k} = {\bf \hat V}_{rv,k-1} \bigcup {\bf \hat v}_{rv,k} $. \\
  5 \textbf{End for } \\
  6 Return $ {\bf \hat V}_{rv} = {\bf \hat V}_{rv,K}$. \\
  \hline
  \end{tabular}
\end{table}

It should be noted that the number of steering vectors in the set ${\bf \hat V}_{rv}$ corresponding to the clutter rank has vital effects on the performance of RTSKA-RD-SA-STAP. To address this problem, we will propose a robust approach in the next section to estimate the clutter rank considering inaccurate prior knowledge.
\begin{table}[!htbp]
  \centering
  \caption{Procedure of the proposed RTSKA-RD-SA-STAP algorithm }\label{CAUP RV InAME-KA}
  \begin{tabular}{l}
  \hline
  \textbf{Initialization:}\\
  ${\bf X} =  [{\bf x}_1, \cdots, {\bf x}_L]$, ${\Delta v_{pm} }$,
  ${\Delta \psi_{m} }$, $v'_p$, ${\psi}'$, $M_e$, $m$. \\

  \textbf{Estimate clutter rank using \eqref{cluRaksubBsubT2}: }\\
  \ $  N_{cx,2} = \sum^{Q}_{q=1} \sum^{K}_{k=1}B^{(q)}_s L^{(q,k)}_{x,\textrm{max}} +1 $.\\

   \textbf{Derive the RD virtual snapshot vector ${\bf z}_r$: }\\
  1 \ {\textrm{Compute}} ${\bf P} $, ${\bf F} $, \textrm{and} ${\bf G} $ {\textrm{using}} \eqref{rdVirSp2},\\
  2 \ ${\bf \hat R} = {\bf X} {\bf X}^H/L$, \\
  3 \ ${\bf z}_r = {\bf G} \textrm{vec}( {\bf \hat R})$. \\
  \textbf{Construct the dictionary ${\bf \Psi}$ as shown in Fig. \ref{dictionary}}. \\
  \textbf{Estimate the clutter subspace ${\bf \hat V}_{rv} $ using Table \ref{cluSpaEst}}.\\
  \textbf{Compute the weight vector using \eqref{filter}}.\\
  \hline
  \end{tabular}
\end{table}

Note that RTSKA-RD-SA-STAP is similar to the InAME-KA \cite{ZYEnhanced2017} but has a key difference. More exactly, in the first stage, prior knowledge is applied to the virtual transformed data in RTSKA-RD-SA-STAP while it is applied to the data received by the ULA in \cite{ZYEnhanced2017}. Hence, RTSKA-RD-SA-STAP has the potential of improved parameter resolution since virtual transformed data has increased DoFs compared with the ULA case in \cite{ZYEnhanced2017}. Furthermore, it should be noted that the formed overcomplete dictionary in this paper is different from that in \cite{ZYEnhanced2017}. In particular, the space-time steering vector set in \cite{ZYEnhanced2017} is of size $MN \times M_d$ and consists of space-time steering vectors corresponding to the ULA. While the overcomplete dictionary ${\bf \Psi}$ in RTSKA-RD-SA-STAP is of size $mN_v \times M_d$ comprised of space-time steering vectors associated with the RD virtual transformed data. Hence, the size and entries of matrix ${\bf \Psi}$ are absolutely different from that of \cite{ZYEnhanced2017}.

\section{Robust Clutter Rank Estimation Considering Inaccurate Prior Knowledge}
The clutter rank is required for estimating the clutter subspace or deriving the eigenalysis-based filter \cite{HaiEig1996}. The well known bandwidth aperture product (termed BT) theorem gives the clutter rank under the continuous aperture and bandwidth scenarios. Specifically, the clutter rank is $BT + 1$, where $B$ is the signal bandwidth and $T$ is the aperture of the sampling array. It has been proved that the earlier Brennan's rule is a special case of the BT theorem \cite{CluttRankGoodman2007}. The continuous aperture means that the signal sampling is discretized with Nyquist sampling interval. The extended BT (EBT) theorem is developed for the sparse aperture and the clutter rank is $\sum^K_{k=1} BT_k + 1$ with $T_k$ being the continuous aperture for the $k$th sub-aperture which is derived by dividing the whole sparse aperture into $K$ sub-apertures \cite{WangSTAP2018,CluttRankGoodman2007}. Both the above mentioned BT theorems are derived under the side-looking array radars and require accurate prior knowledge such as platform velocity, and crab angle etc. Here, we propose a rule for clutter rank estimation under some practical error factors. The rule is derived from the asymptotic expansion of the EBT theorem.

In ideal cases, assuming the spatial and Doppler frequency $\vartheta^{(q)}$ and $\varpi^{(q)}$ for the direction angle ${\theta}^{(q)}$, the component in space-time steering vector induced by the $n$th sensor and the $m$th pulse is given by
\begin{eqnarray}\label{steVnmq}
\begin{split}
  {\bf v}^{(q)}_{n,m}
    = e^{j2\pi [ \vartheta^{(q)} d_{(n-1)}  + \varpi^{(q)} t_{(m-1)}] },
\end{split}
\end{eqnarray}
where $d_{(n-1)}$ denotes the $n$th sensor positions with respect to the first sensor, and $t_{(m-1)}$ is the time instant of the $m$th pulse. Assuming the ratio $\beta^{(q)} = \frac{\varpi^{(q)}}{\vartheta^{(q)}} ( \vartheta^{(q)} \neq 0)$, \eqref{steVnmq} can be simplified as
\begin{eqnarray}\label{steVnmqSim}
\begin{split}
  {\bf v}^{(q)}_{n,m}
    = e^{j2\pi f^{(q)}_s [d_{(n-1)}  + \beta^{(q)} t_{(m-1)}] }.
\end{split}
\end{eqnarray}
Here, $f^{(q)}_s = \vartheta^{(q)}$ denotes the spatial frequency associated with an equivalent sampling array with array sensors located at $d_{(n-1)}  + \beta^{(q)} t_{(m-1)},n=1,\cdots, N, m=1,\cdots, M$. Then, for continuous signals, the $i$th row and the $j$th column component of the clutter covariance matrix is given by
\begin{eqnarray}\label{covEquviV}
\begin{split}
   r_c(p^{(q)}_{i} - p^{(q)}_{j} )  = \int_G P_c(f^{(q)}_s) e^ { j2 \pi f^{(q)}_s (p^{(q)}_i- p^{(q)}_j ) } d{f^{(q)}_s },
\end{split}
\end{eqnarray}
where $ P_c(f^{(q)}_s)$ is the power spectrum of the received signal, $G$ is the range of spatial frequencies $f^{(q)}_s $, and $p^{(q)}_{i}$ and $p^{(q)}_{j}$ are the $i$th and $j$th sensor position of the equivalent array. For a ULA and a fixed PRI, the sensor positions are $p^{(q)} = (n-1) + \beta^{(q)} (m-1)$, $n=1,\cdots,N$, $m=1,\cdots,M$. By performing eigenvalue decomposition on the clutter covariance matrix, the eigenvalue and eigenvector have the following relationship
\begin{eqnarray}\label{eigencovEquviV}
\begin{split}
  \lambda_{i} u_i(p_{x}) = \int_0^{L_x} r_c(p_{x} - q_{x}) u_i(q_{x}) d_{q_{x}},
\end{split}
\end{eqnarray}
where $\lambda_i $ and $u_i(p_{x})$ are the $i$th eigenvalue and the corresponding eigenvector, $p_{x}$ and $q_{x} $ denote the positions of the equivalent array, and $L_x$ stands for the aperture of the equivalent array. It is known that $r_c(p_{x}) = P \textrm{sinc} (B_s p_{x})$ when the power of the signal is uniformly distributed with the power spectrum density being $P$ and the bandwidth being $B_s$. Here $\textrm{sinc}(x) = \sin(\pi x) / (\pi x)$. Thus, the clutter rank of \eqref{eigencovEquviV} is $N_{cx} = B_sL_x +1$ \cite{CluttRankGoodman2007}.

In practical applications, due to the presence of imperfections, the ideal platform velocity $v_p$ and crab angle $\psi$ are unavailable and the actual Doppler frequency corresponding to a certain direction angle is uncertain. We thus propose to estimate the Doppler frequency as a set of Doppler frequency points. Particularly, for a spatial frequency ${\vartheta}^{(q)}$ corresponding to the given direction angle ${\theta}^{(q)}$, a set of ratios $\{ \beta^{(q)}_i \}^{M_e}_{i=1}$ and the corresponding equivalent arrays can be directly determined. To take such uncertainty into account, an equivalent array with the largest aperture is chosen to determine the clutter rank for a single given direction angle. Hence, the clutter rank for the direction angle ${\theta}^{(q)}$ can be estimated as
\begin{eqnarray}\label{cluRaksubBsubT1}
\begin{split}
  N^{(q)}_{cx} = B^{(q)}_s \textrm{max}( \{ L^{(q)}_{x,i} \}^{M_e}_{i=1}) +1,
\end{split}
\end{eqnarray}
where $B^{(q)}_s$ denotes the signal bandwidth of $q$th spatial frequency signal, $L^{(q)}_{x,i}$ is the aperture of the $i$th equivalent array with array positions $p^{(q)}_i = d_{(n-1)} + \beta^{(q)}_i t_{(m-1)}$, $n=1,\cdots,N$ and $m=1,\cdots,M$, and $\textrm{max}(\cdot)$ returns the maximum value of the argument.

Letting $Q$ be the number of sampled direction angles at the given range bin, one can get the corresponding clutter rank as
\begin{eqnarray}\label{cluRaksubBsubT1}
\begin{split}
  N_{cx,1} = \sum^{Q}_{q=1}B^{(q)}_s \textrm{max}( \{ L^{(q)}_{x,i} \}^{M_e}_{i=1}) +1.
\end{split}
\end{eqnarray}
It should be noted that for practical implementations, $B^{(q)}_s$ can be approximately equal to the difference between the $(q-1)$th and $q$th spatial frequency, and $L^{(q)}_{x,i} =  (N-1) + \beta^{(q)}_i (M-1)$ when the ULA and fixed pulses interval structure satisfy the Nyquist sampling condition. On the other hand, we notice that the above derivation is based on the fact that $\beta^{(q)} $ is nonzero, whereas this constant may be disturbed when $f^{(q)}_s = 0$ for direction angle $\theta = 0^{\circ}$. In this case, the equivalent array cannot be achieved. To proceed, we approximate the signal by dropping the spatial frequency and the corresponding Doppler frequency points for direction angle $\theta = 0^{\circ}$. Alternatively, one may also sample the direction angle onto even samples such that $f^{(q)}_s$ is nonzero.

Moreover, when the equivalent array is a sparse array, by dividing the sparse array into multiple continuous sub-arrays to satisfy the Nyquist sampling condition, the extended expression for clutter rank estimation is given by
\begin{eqnarray}\label{cluRaksubBsubT2}
\begin{split}
  N_{cx,2} = \sum^{Q}_{q=1} \sum^{K}_{k=1}B^{(q)}_s L^{(q,k)}_{x,\textrm{max}} +1,
\end{split}
\end{eqnarray}
where $K$ is the total number of sub-arrays, and $L^{(q,k)}_{x,\textrm{max}}$ is the aperture of the $k$th sub-array of the $q$th equivalent sparse array with the maximum aperture, i.e., $L^{(q,k)}_{x,\textrm{max}} = \{ \textrm{max}( \{ L^{(q)}_{x,i} \}^{M_e}_{i=1}) \}^k $. Finally, the proposed robust clutter rank estimation approach is summarized in Table \ref{cluRanEst}.
\begin{table}[!htbp]
  \centering
  \caption{Proposed clutter rank estimation approach }\label{cluRanEst}
  \begin{tabular}{l}
  \hline
  \\[-2.5mm]
  Step 1: Determine the spatial frequency bandwidth $ \{ B^{(q)}_s \}^Q_{q=1} $  \\
   \quad \quad \ \quad using $\{f^{(q)}_s \}^Q_{q=1}$. \\
  Step 2: Compute the equivalent array positions $p^{(q)}$ as \\
   \quad \quad \ \quad  $p^{(q)} = d_{(n-1)} + \beta^{(q)} t_{(m-1)}$, \\
   \quad \quad \ \quad  $n=1,\cdots,N$, $m=1,\cdots,M$ for each $f^{(q)}_s$.\\
  Step 3: Compute the equivalent aperture set $\{ L^{(q)}_{x,i} \}^{M_e}_{i=1}$ for each $p^{(q)}$.\\
  Step 4: Divide the equivalent aperture into $K$ parts as $L^{(k)}_{x}$, \\
   \quad \quad \ \quad  $k=1,\cdots,K$ according to the Nyquist sampling interval. \\
  Step 5: Estimate the clutter rank using \eqref{cluRaksubBsubT2} as \\
   \quad \quad \ \quad $N_{cx,2} = \sum^{Q}_{q=1} \sum^{K}_{k=1}B^{(q)}_s L^{(q,k)}_{x,\textrm{max}} +1$.
  \\ [1mm]
 \hline
\end{tabular}
\end{table}

Interestingly, it can be noticed that if the bandwidth is continuous and the equivalent array aperture is also continuous or satisfies the Nyquist sampling condition, then we have $N_{cx} = B_s L_{x} +1$, which is the standard BT theorem. Furthermore, if the bandwidth is continuous while the equivalent array aperture has holes, then we have  $N_{cx} = \sum^{K}_{k=1} B_s L^{(k)}_{x} +1$, which is just the EBT theory. Therefore, \eqref{cluRaksubBsubT2} is a generalized form of \eqref{cluRaksubBsubT1}, and the proposed approach reduces to that in our previous work \cite{WangSTAP2018}. This implies that the proposed rule for clutter rank estimation generalizes the traditional ones to more practical scenarios and is applicable to cases including uniform, nonuniform sampling and inaccurate aircraft velocity and crab errors.

\section{ Analysis of RTSKA-RD-SA-STAP}  \label{analysis}
In this section, we first analyze the convergence of the constructed virtual snapshot by deriving a closed form expression of its estimation error distribution. Then, implementations of the proposed RTSKA-RD-SA-STAP algorithm is discussed followed by the computational complexity analysis.

\subsection{ Convergence Analysis of Virtual Construction }
Since the covariance matrix ${\bf R}$ is usually replaced by its estimated counterpart ${\bf \hat R}$ in practice, i.e., the constructed virtual snapshot is obtained from the covariance matrix estimate ${\bf \hat R}$ in \eqref{CovHat}. Hence, the practical resultant virtual snapshot may be deviated from the actual virtual snapshot. In order to observe the changes of the constructed virtual snapshot along the increase of the number of training samples, we derive the estimation error distribution of the constructed virtual snapshot.

Note that the virtual snapshot ${\bf z}_r$ in \eqref{rdVirSp1} can be equivalently rewritten as
\begin{eqnarray}\label{Anas1_rdVirSp1}
\begin{split}
  {\bf z}_r
  & = \left[
     \begin{matrix}
      {\bf P} ({\bf u}^T(\varpi_1) \otimes {\bf u}^H(\varpi_1 )) {\bf r} \\
      {\bf P} ({\bf u}^T(\varpi_2) \otimes {\bf u}^H(\varpi_2 )) {\bf r} \\
     \cdots\\
      {\bf P} ({\bf u}^T(\varpi_m) \otimes {\bf u}^H(\varpi_m )) {\bf r} \\
     \end{matrix}
     \right]   = \bf D {\bf r},
\end{split}
\end{eqnarray}
where
\begin{eqnarray}\label{Anas2_rdVirSpM}
\begin{split}
   \bf D = \left[
     \begin{matrix}
      {\bf P} ({\bf u}^T(\varpi_1) \otimes {\bf u}^H(\varpi_1 ))  \\
      {\bf P} ({\bf u}^T(\varpi_2) \otimes {\bf u}^H(\varpi_2 ))  \\
     \cdots\\
      {\bf P} ({\bf u}^T(\varpi_m) \otimes {\bf u}^H(\varpi_m ))  \\
     \end{matrix}
     \right].
\end{split}
\end{eqnarray}
As mentioned earlier, in practice the ideal covariance matrix $\bf R$ is unavailable and is estimated with a finite number of training samples as in \eqref{CovHat}. Hence, $\bf z_r$ and $\bf r$ in \eqref{Anas1_rdVirSp1} should be replaced by $\bf \hat z_r$ and $\bf \hat r$, which are respectively the virtual snapshot estimate and the vectorized form of the covariance matrix estimate $\bf \hat R$, i.e., $\bf \hat r = \textrm{vec}(\bf \hat R)$ and
\begin{eqnarray}\label{rdVirSpHat}
\begin{split}
    {\bf \hat z}_{ r } = {\bf D} { \bf \hat r }.
\end{split}
\end{eqnarray}
In order to analyze the estimation error of the virtual snapshot in \eqref{rdVirSpHat}, first, note that the estimation error $\bf \hat r - \bf r$ obeys an asymptotic zero-mean normal distribution \cite{BCovarian1998}, i.e.,
\begin{eqnarray}\label{biasDistr}
\begin{split}
    {\bf \hat r} - { \bf r } \thicksim As \mathcal{N}(0,\frac{1}{L} ({\bf R}^T \otimes {\bf R})),
\end{split}
\end{eqnarray}
where $As \mathcal{N} (\mu, \boldsymbol \Sigma)$ represents the asymptotic normal distribution with mean $\mu$ and covariance matrix $\boldsymbol \Sigma$.

According to \eqref{rdVirSpHat} and \eqref{biasDistr}, we know that the estimation error of the virtual snapshot estimate $\bf \hat z_r$ obeys an asymptotic standard normal distribution, i.e.,
\begin{eqnarray}\label{biasDistr1}
\begin{split}
    {\bf \hat z}_{r }- { \bf z }_{r} \thicksim As \mathcal{N}(0,\bf C),
\end{split}
\end{eqnarray}
where $\bf C$ is given by
\begin{eqnarray}\label{biasDistr1Matrix}
\begin{split}
 {\bf C} = \frac{1}{L} {\bf D} ({\bf R}^T \otimes {\bf R}){\bf D}^H.
\end{split}
\end{eqnarray}
According to \eqref{biasDistr1} and using ${\bf {\tilde{z}}}_{r} = {\bf \Psi} {\bf \tilde{p}} + {\sigma}_n^2 {\bf G} {\bf e}_0$, we have
\begin{eqnarray}\label{biasDistr1Matrix1}
\begin{split}
    {\bf C }^{-\frac{1}{2}} ( {\bf \hat{\tilde{z}}}_{r} - {\bf \Psi} {\bf \tilde{p}}- {\sigma}_n^2 {\bf G} {\bf e}_0 ) \thicksim As \mathcal{N}(0,\bf I),
\end{split}
\end{eqnarray}
where we have $ [ {\bf C }^{-\frac{1}{2}} ( {\bf \hat{\tilde{z}}}_{r} - {\bf \Psi} {\bf \tilde{p}}- {\sigma}_n^2 {\bf G} {\bf e}_0 ) ]_i \thicksim As \mathcal{N}(0,1)$, for $i=1,\cdots,(mN_v)^2$, and $[\cdot]_i$ denotes the $i$th entry of a vector. Moreover, by using the fact that the distribution of a sum of the squares of $N$ independent standard normal random variables obeys the Chi-square distribution with $N$ DoFs, we have
\begin{eqnarray}\label{biasDistr1Matrix2}
\begin{split}
    & \| {\bf C }^{-\frac{1}{2}} ( {\bf \hat{\tilde{z}}}_{r} - {\bf \Psi} {\bf \tilde{p}} - {\sigma}_n^2 {\bf G} {\bf e}_0 ) \|^2_2 = \\
    &\sum^{(M_rN_v)^2}_{i=1} | [ {\bf C }^{-\frac{1}{2}} ( {\bf \hat{\tilde{z}}}_{r} - {\bf \Psi} {\bf \tilde{p}} - {\sigma}_n^2 {\bf G} {\bf e}_0 ) ]_i|^2 \thicksim As \mathcal{\chi}^2(m^2N^2_v),
\end{split}
\end{eqnarray}
where $As \mathcal{\chi}^2(m^2N^2_v)$ stands for the asymptotic Chi-square distribution with $m^2N^2_v$ DoFs. The Chi-squared distribution converges to a normal distribution with approximate mean of $m^2N^2_v$ when the number of training samples is close to infinity.

It is worth noting that the above derivation is based on the assumption that the matrices $\bf C$ and $\bf R$ are exactly known. However, $\bf R$ can only be estimated from a set of finite training samples and hence, the covariance matrix $\bf C$ should be estimated based on $\bf \hat R$ as
\begin{equation}\label{biasDistrHat}
\begin{split}
   {\bf \hat C} \triangleq \frac{1}{L} {\bf D} ({\bf \hat R}^T \otimes {\bf \hat R}){\bf D}^H.
\end{split}
\end{equation}
Hence, the covariance matrix $\bf C$ in \eqref{biasDistr1}, \eqref{biasDistr1Matrix1}, and \eqref{biasDistr1Matrix2} is replaced by its estimate $ {\bf \hat C}$ in \eqref{biasDistrHat} in real systems.

It can be seen that the variance of the virtual snapshot estimate in \eqref{biasDistr1Matrix2} depends on the number of training samples $L$, covariance matrix ${\bf R}$ (or clutter-to-noise ratio (CNR)), and the DoFs of the virtual signal $m^2N^2_v$. More exactly, the estimation variance increases exponentially with the increase of CNR, decreases linearly with the increase of number of training samples $L$, and is close to a nonzero value when the number of training samples $L$ goes to infinity. A similar issue is also reported in problems of DOA estimation or spatial beamforming for virtual signals, such as \cite{MWang2018} and \cite{PPal2010}. Therefore, compared with the conventional STAP approaches, the STAP algorithm based on the virtual array/samples requires a greater number of training samples. This could be viewed as a method that uses training samples from the range cell to compensate for the loss in spatial samples. Additionally, it should be noted that the convergence analysis of the virtual construction is related to the convergence of the proposed algorithm. The convergence of the proposed algorithm not only depends on the estimation errors of the virtual snapshot but also on the selected sparse recovery algorithm. This remains an open issue, which will be dealt with in our future work. In this work, we rely on numerical simulations to evaluate the convergence performance of the proposed algorithm. As can be seen from the simulation results, a small number of training samples is needed in RTSKA-RD-SA-STAP to achieve satisfactory performance. This is because the proposed RTSKA-RD-SA-STAP algorithm has exploited the prior knowledge and the sparsity of the clutter.

\subsection{ Implementation Discussions }
As suggested in \cite{Knowledge2006,GuerciKA2006} and \cite{ZYEnhanced2017}, the platform velocity and crab angle may fluctuate over a range due to practical imperfections such as those from the environments and aircraft controls and thus be time-variant quantities. This implies that the covariances for each range bin can be different from each other during a CPI. In the knowledge-aided STAP algorithm \cite{Knowledge2006}, it is required to form range-varying covariances for instantaneously tracking the variations in clutter environments, resulting in a high computational burden. However, in the proposed RTSKA-RD-SA-STAP algorithm, it is noted that the formulation in \eqref{difTimFreq2} is derived based on the setting of range values of platform velocity and crab angle errors as well as its measured values of platform velocity and crab angle. As illustrated subsequently in simulation results,  RTSKA-RD-SA-STAP is robust to measured errors and shows good performance for certain error range values, where the clutter subspace is only estimated once for all range bins adjacent to the CUT. Therefore, RTSKA-RD-SA-STAP is easily implemented and has low complexity as compared to the knowledge-aided STAP algorithm in \cite{Knowledge2006}. Similar to \cite{ZYEnhanced2017}, RTSKA-RD-SA-STAP can also be applied to multiple consecutive CPIs, provided that there is a judicious setting of prior knowledge.

\begin{figure}[!htbp]
\centering
  \includegraphics[width=88mm]{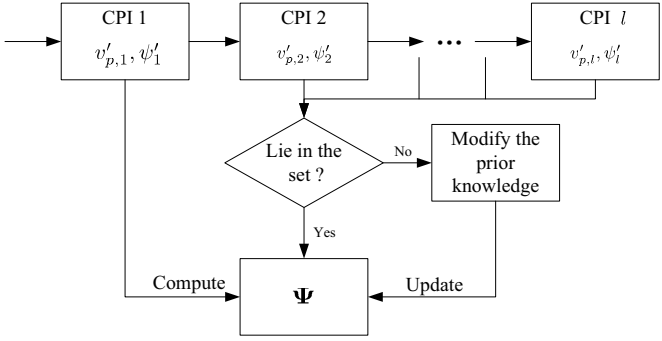}
  \caption{\small The implementation of RTSKA-RD-SA-STAP in case of multiple consecutive CPIs. } \label{fig:ReceivedCPIs}
\end{figure}

For instance, there are $l$ CPIs as shown in Fig. \ref{fig:ReceivedCPIs}, where $v'_{p,j},j=1,\cdots,l $ denotes the measured platform velocity which lies within a certain range, and $\psi'_j,j=1,\cdot,l $ denotes the measured crab angle which also lies within a certain range. Given the true platform velocity $v_{p,1}$ and crab angle $\psi_{1}$ in the first CPI, if $v'_{p,j} \in [v_{p,1} -\Delta v_{pm},v_{p,1} +\Delta v_{pm}] $ and $\psi'_{j} \in [\psi_{1} -\Delta \psi_{m},\psi_{1} +\Delta \psi_{m}]$ for $j=1,\cdots,l$, where $\Delta v_{pm}$ and $\Delta \psi_{m}$ are prior knowledge of error range values of platform velocity and crab angle, respectively, we only compute the dictionary once during the $l$ CPIs. Meanwhile, if $v'_{p,j}$ or $\psi'_{j}$ $(1 \leq j \leq l)$ does not lie in that set, we modify the prior knowledge and update the dictionary. In fact, it is found experimentally that prior knowledge errors $\Delta v_{pm}$ or $\Delta \psi_{m}$ can be chosen from a relatively wide range such that $v'_{p,j}$ or $\psi'_j $ $(1 \leq j \leq l)$ lie in those sets, without significantly affecting the performance. This implies that the dictionary in  RTSKA-RD-SA-STAP is still constructed only once with a certain range of prior knowledge, which is suitable for real systems.

\subsection{Complexity Analysis}
The computational complexity of the proposed RTSKA-RD-SA-STAP algorithm mainly depends on the complexity of computing ${\bf \hat R}$ in \eqref{CovHat} and estimating the clutter subspace by the OMP-like method. Thus, the overall complexity is $\mathcal{O} (L (NM)^2 + ( k_x m N_{v}  N'_c M_e) )$, where $k_{x}$ is the clutter rank estimation value. For comparisons, the complexity of the mDT\cite{YLWVari2000}, JDL \cite{HWOnAdp1994}, PC \cite{HaiEig1996}, InAME-KA \cite{ZYEnhanced2017}, Virtual-Smoothed-PC  (termed VS-PC) \cite{WangSTAP2018}, and FD-SA-STAP algorithm \cite{wxySparse2018} are also presented in Table \ref{Complexity}, where $N_L$ and $M_L$ are the number of localized angle and Doppler channels, $k$ is the number of iterations in the InAME-KA algorithm, and $N_{vs}$ and $M_{vs}$ are the number of sensors and pulses after virtual smoothing process, respectively. Compared with the FD-SA-STAP algorithm, the proposed RTSKA-RD-SA-STAP algorithm incorporates a two-stage RD process, which results in a dimensionality reduced optimization problem and does not require the covariance matrix inversion. Therefore, it can achieve a great complexity reduction. However, it requires more computational operations than the mDT, JDL, PC, and InAME-KA, since it operates on the virtual signal. Fortunately, RTSKA-RD-SA-STAP can provide much better performance than these algorithms as shown in the following section.

\section{Numerical Simulations} \label{Simulation}
In this section, various simulation results are provided to validate the theoretical derivation and to demonstrate the performance of RTSKA-RD-SA-STAP. Radar parameters are assumed that $h_p =$ 125m/s, $v_p=$ 4000m, $T_r=1/4000$s, and $d_0=0.0625$m. The clutter in a given range bin is divided into $N_c=$ 361 patches and each patch is assumed to be IID and  be distributed as the zero mean complex Gaussian process with variance equal to $10^{10/(361 \textrm{CNR})}$ for a given $\textrm{CNR}$ in decibel scale. The noise at the receiver is drawn from a zero mean complex Gaussian process with variance $\sigma^2_n=1$. In our examples, unless otherwise stated, all results are calculated by averaging the results over 500 Monte Carlo experiments.

\begin{table}[!htbp]
\caption{Computational complexity comparisons}
\label{Complexity}
\centering
\begin{tabular}{|c|c|c|}
\hline
\multirow{2}*{Algorithms}  & \multicolumn{2}{|c|} {Computational complexity for computing}\\
\cline{2-3}
             & covariance matrix  &  filter weight\\
 \hline
 mDT \cite{YLWVari2000}    & $ \mathcal{O} (mLN^2M )$         & $ \mathcal{O}[ (mN)^3] $ \\
\hline
 JDL \cite{HWOnAdp1994}  & $ \mathcal{O} (L N_{L}M_{L} NM )$  & $ \mathcal{O} [ (N_{L}M_{L})^3 ] $  \\
\hline
 PC \cite{HaiEig1996}    & $ \mathcal{O} [L(NM)^2]$           & $ \mathcal{O} [(NM)^3]$  \\
\hline
 InAME-KA \cite{ZYEnhanced2017} & $ \mathcal{O} (kNM N'_c M_e)$     & $\mathcal{O} (k(NM)^2)$\\
 \hline
 VS-PC \cite{WangSTAP2018}  & $ \mathcal{O} (L (NM)^2)$   & $ \mathcal{O} [ ( N_{vs}M_{vs})^3 ]$ \\
 \hline
 FD-SA-STAP \cite{wxySparse2018} & $ \mathcal{O} [ (N_d N_s)^3]$ & $\mathcal{O} [ (N_{v} M_{v})^3]$ \\
 \hline
 RTSKA-RD-SA-STAP       & $ \mathcal{O} (L (NM)^2)$     & $ \mathcal{O} ( k_{x} m N_{v} N'_c M_e)$\\
 \hline
\end{tabular}
\end{table}

\subsection{Clutter Rank Estimation}
In this subsection, we assess the accuracy of the proposed clutter rank estimation approach (marked with solid line) in various scenarios. More precisely, the CNR is 40dB, four cases (i.e., $\beta$ = 0.6, 1 with $\psi = 0^{\circ}$, and $\beta$ = 0.6, 1 with $\psi = 90^{\circ}$) are considered for two types of radar sampling configurations (i.e., ULA and CPA). For the ULA radar, the number of sensors is $N=10$ and the number of pulses in one CPI is $M=10$, while for the CPA radar, the number of sensors is still 10 with coprime factors $N_1=3$ and $N_2=5$, the number of pulses is the same as that in the ULA radar. As a comparison, the clutter rank estimates of the BT theorem for the ULA radar (marked with $\textrm{o}$) and the method of \cite{WangSTAP2018} for the CPA radar (marked with $\times$) are shown for performance evaluation.

\begin{figure}[!htbp]
\centering
  \subfigure[ULA for an ideal case]{\label{fig1}
  \includegraphics[width=42mm]{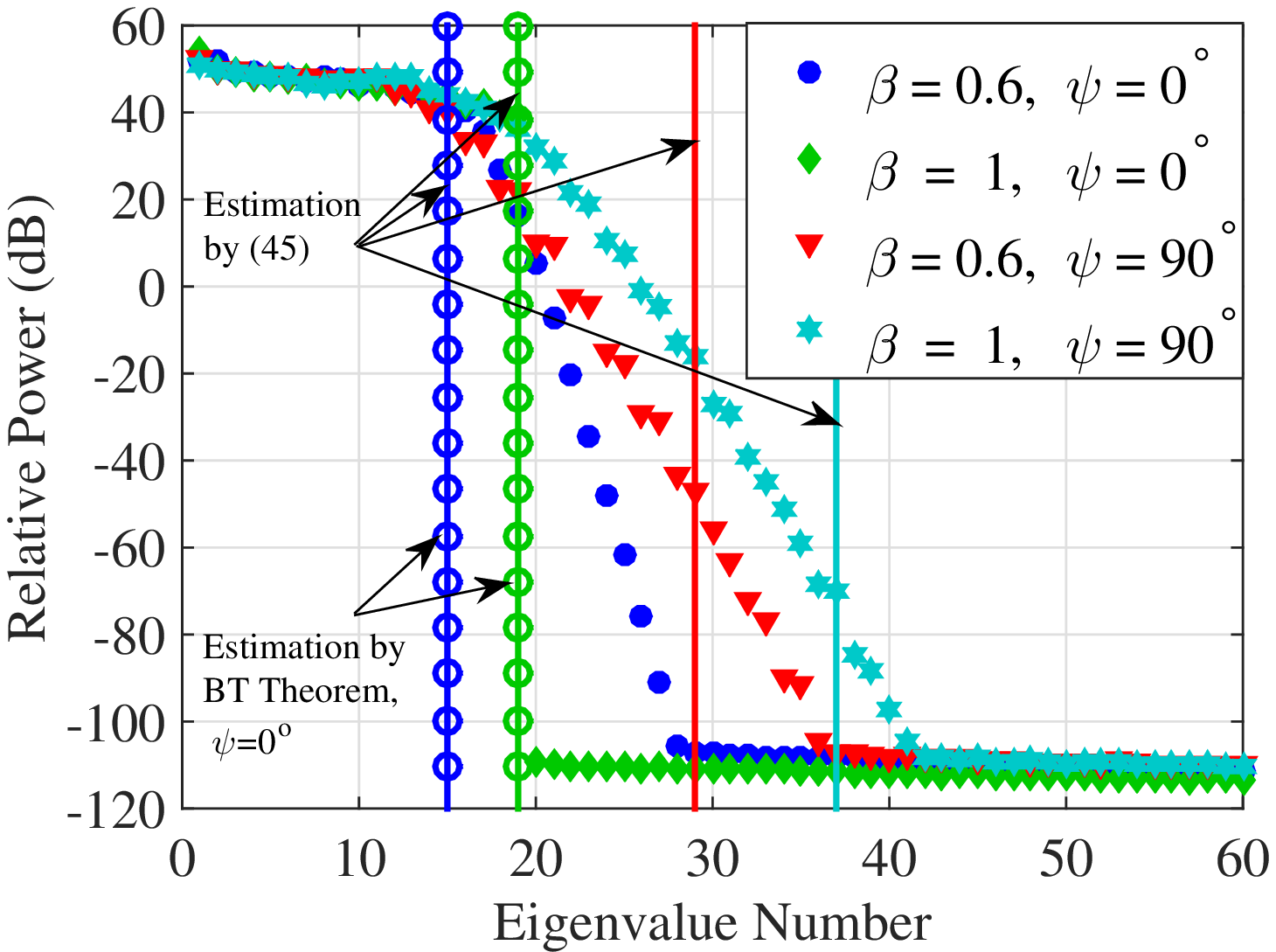}}
  \subfigure[CPA for an ideal case]{\label{fig2}
  \includegraphics[width=42mm]{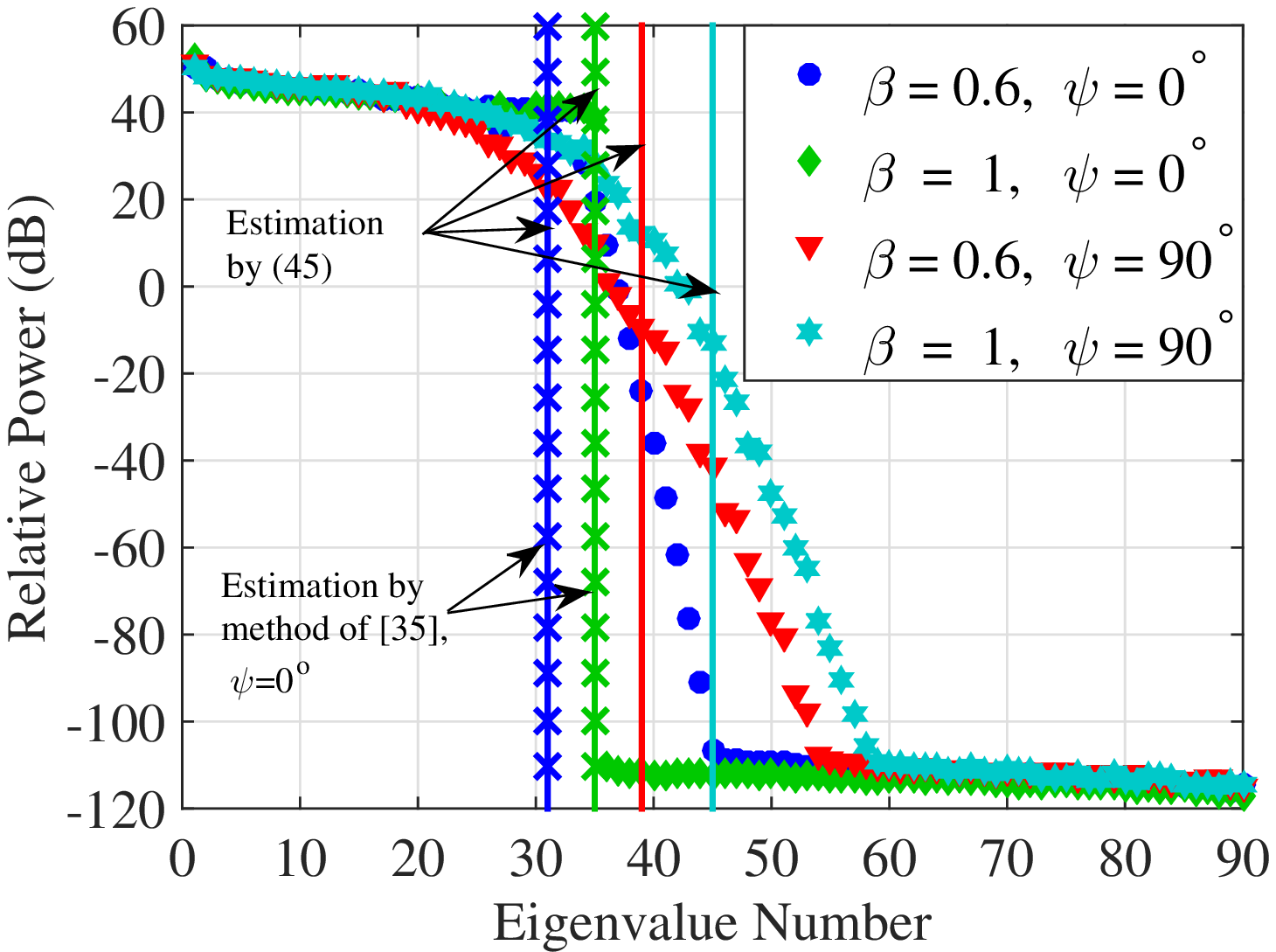}}
  \subfigure[ULA with prior knowledge errors]{\label{fig3}
  \includegraphics[width=42mm]{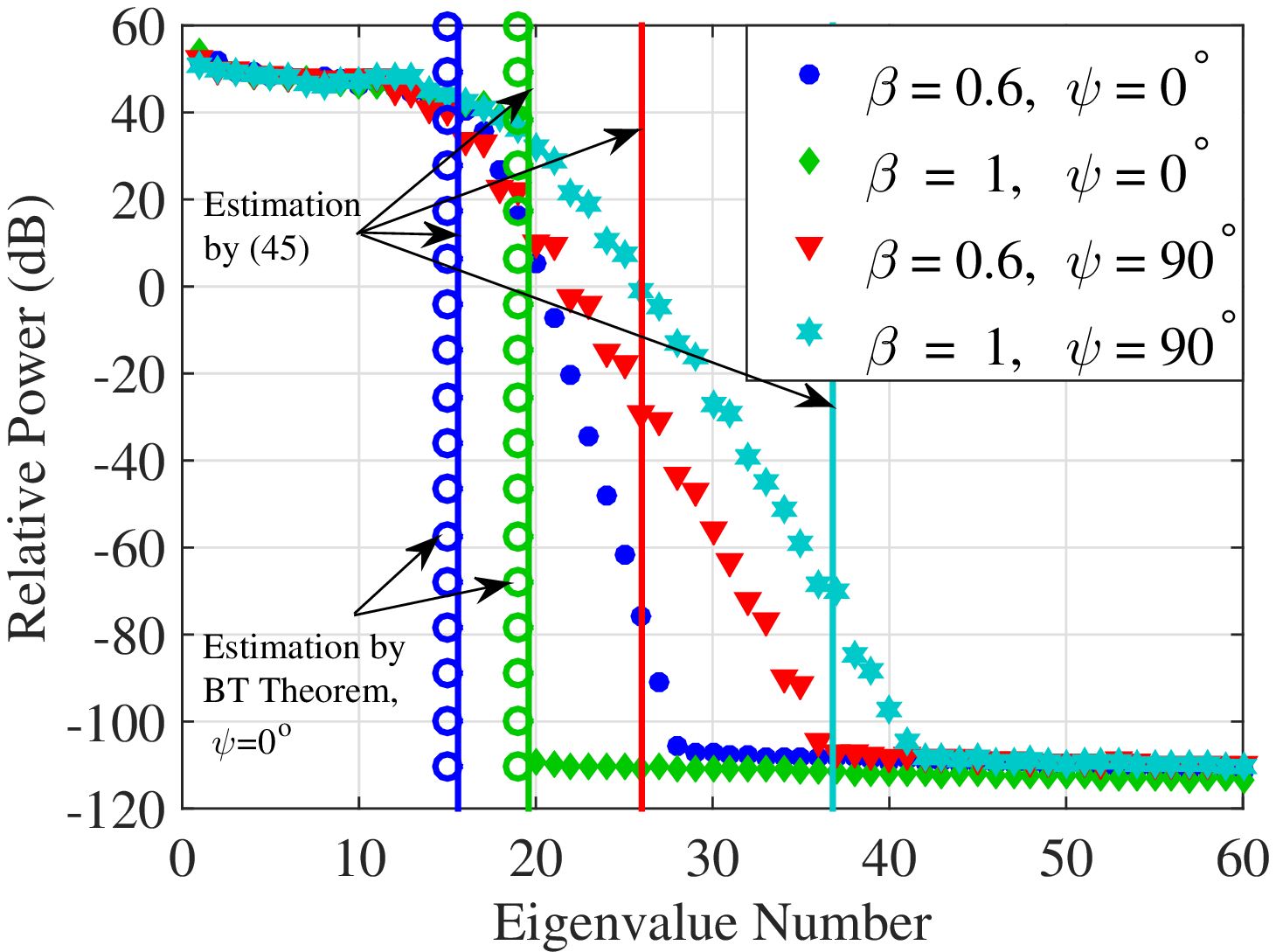}}
  \subfigure[CPA with prior knowledge errors]{\label{fig4}
  \includegraphics[width=42mm]{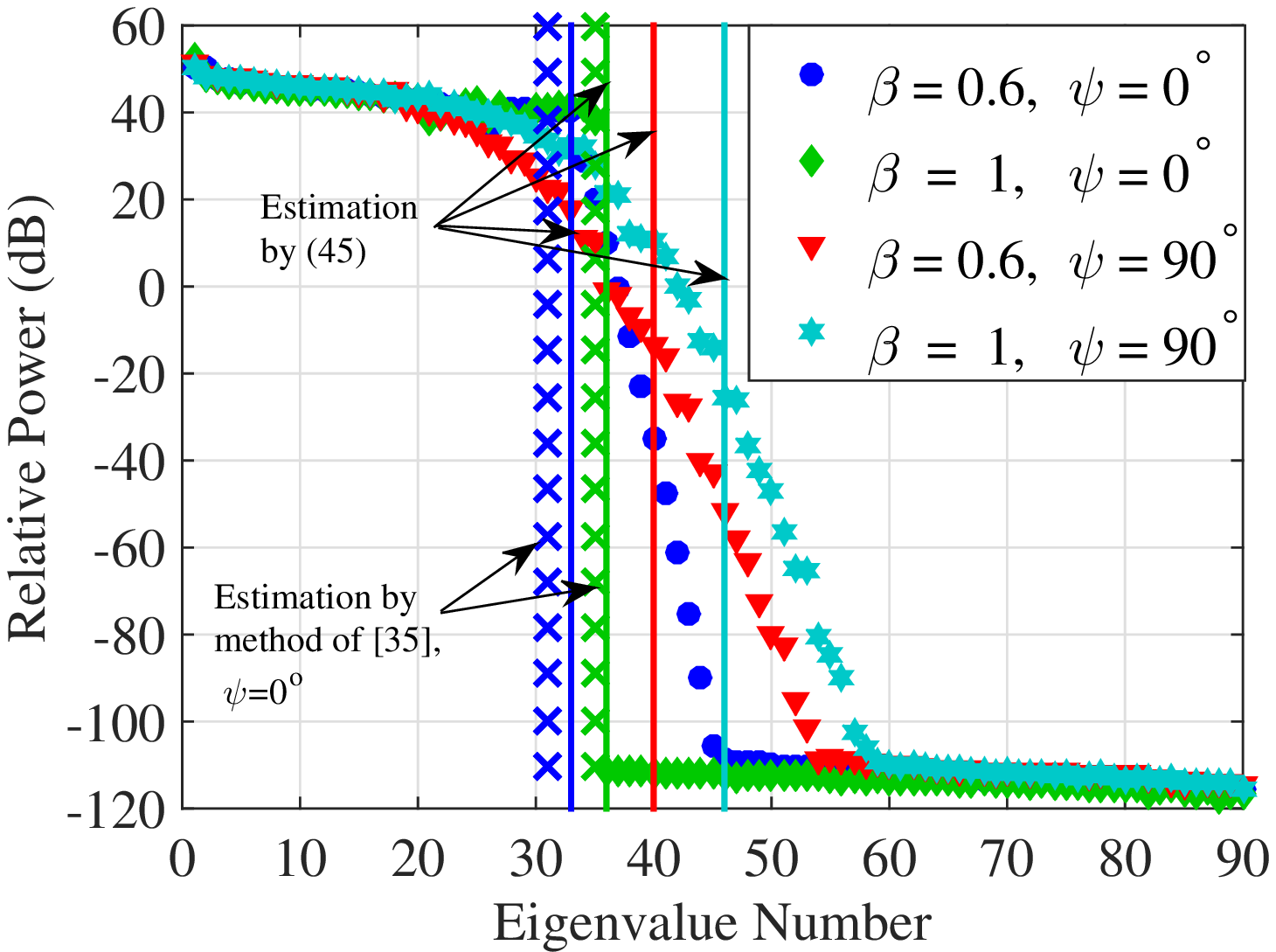}}
  \caption{\small Clutter rank estimation results of the proposed approach (marked with solid line), the BT theorem (marked with $\textrm{o}$), and the method in \cite{WangSTAP2018} (marked with $\times$). }\label{fig:clutRak1}
\end{figure}

Fig. \ref{fig:clutRak1} shows the resultant clutter rank estimates using different approaches. The proposed clutter rank estimation approach is used for both side-looking and non-side-looking radar cases with $ 0 ^{\circ} \leq \psi \leq 90^{\circ}$ while both the BT theorem and the method of [35] are only exploited for side-looking case with $\psi = 0^{\circ}$. It can be seen that the results of the proposed approach for side-looking case with $\psi = 0^{\circ}$ are the same as that of the BT theorem for the ULA radar and as that of method of [35] for the CPA radar with $\psi =0^{\circ}$ in ideal cases (i.e. without prior knowledge errors). Again, for a number of simulations of various prior knowledge errors (not shown), it is found that neither the BT theorem nor the method \cite{WangSTAP2018} can provide accurate clutter rank estimates for side-looking cases with $\psi = 0^{\circ}$ when some error occurs (assumed that $\Delta v_{pm} =5 \textrm{m/s}$, $\Delta v^{\prime}_{pm}= 0.5 \Delta v_{pm} $, $\Delta \psi_{m} =4^{\circ}$, and $\Delta \psi^{\prime}_{m} = 0.5 \Delta  \psi_{m}$). As expected, the proposed approach offers satisfactory performance for side-looking case with $\psi = 0^{\circ}$ since the prior knowledge of errors of platform velocity and crab angle is considered in the proposed approach. Furthermore, for the non-side-looking case (i.e. $ 0^{\circ} \leq \psi \leq 90^{\circ}$), the BT theorem and the method of [35] are not applicable since the crab angle is nonzero. However, as can be seen from Fig. \ref{fig:clutRak1}, the proposed approach is still applicable and provides satisfactory clutter rank estimates. Note that we only plot the representative cases with $\psi=90^{\circ}$ due to space limitation as the number of simulations of various values of $\psi$ share similar results. These results indicate that the proposed approach can not only provide good results for both side-looking and non-side looking radars but also with prior knowledge in the presence of errors, and hence it is a more effective approach as compared to the method of \cite{WangSTAP2018}.

\subsection{Virtual Snapshot Estimation Error Distribution of \eqref{biasDistrHat} }
\begin{figure}[!htbp]
\centering
  \subfigure[]{\label{fig1err}
  \includegraphics[width=85mm]{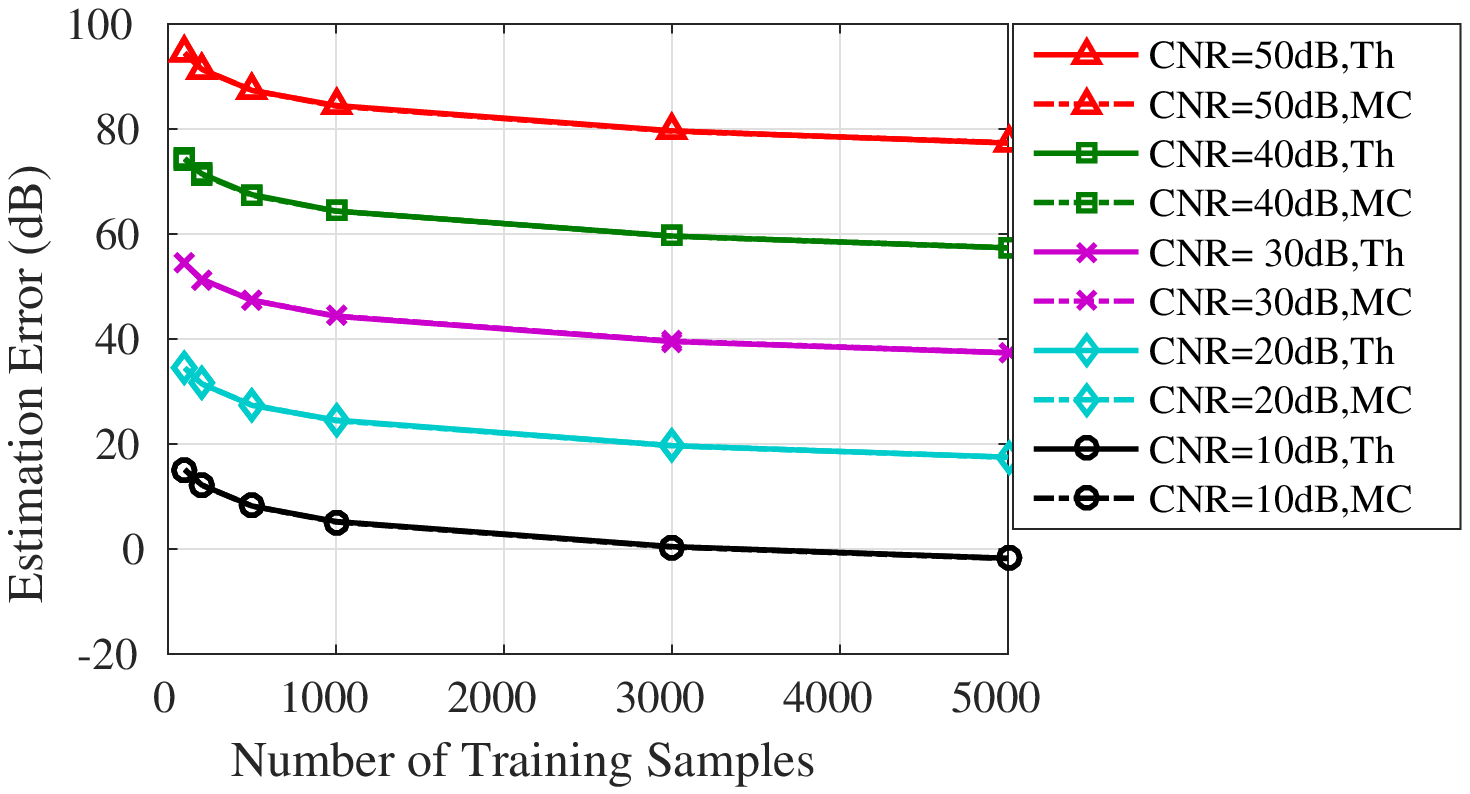}}
  \subfigure[]{\label{fig2err}
  \includegraphics[width=85mm]{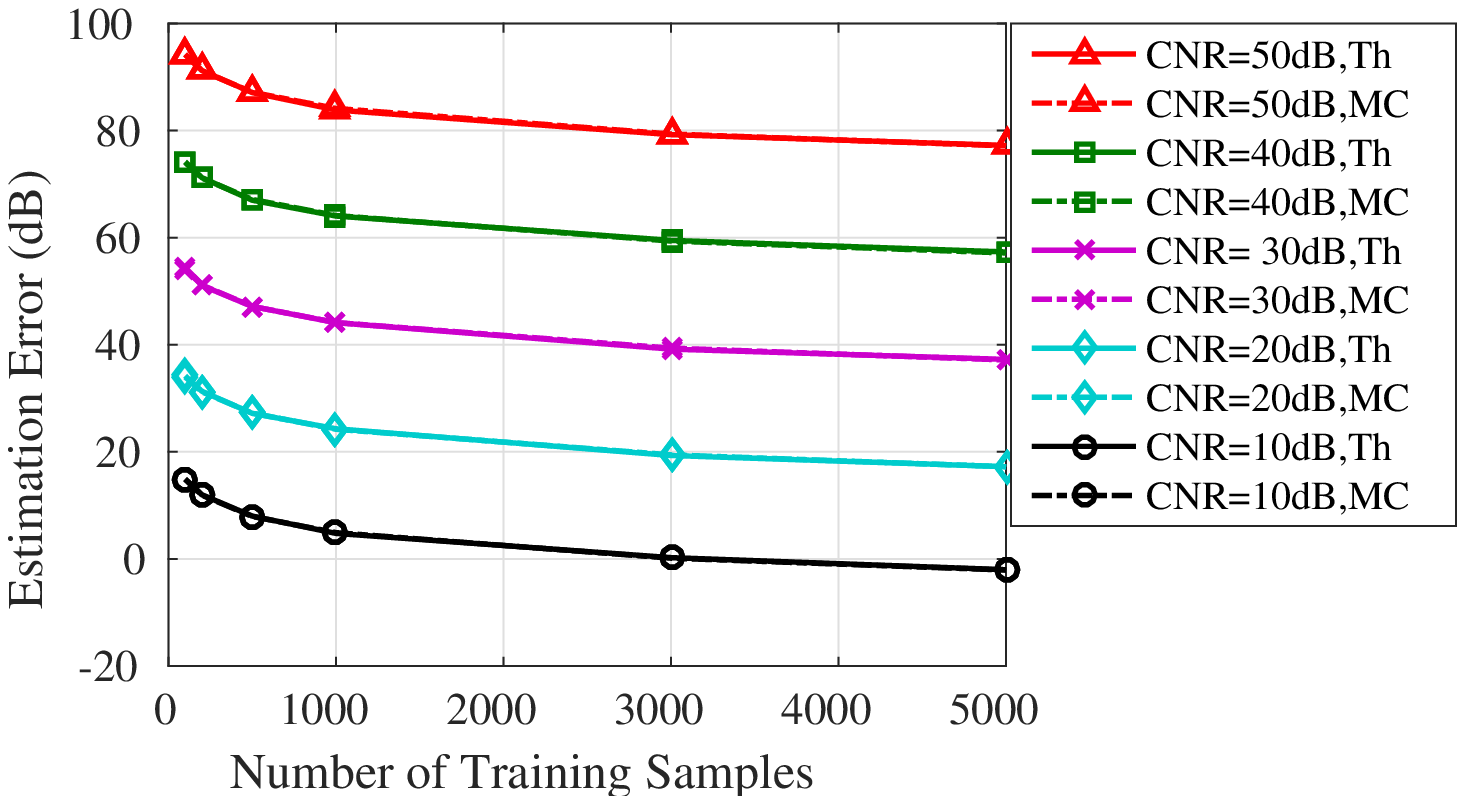}}
  \subfigure[]{\label{fig3err}
  \includegraphics[width=85mm]{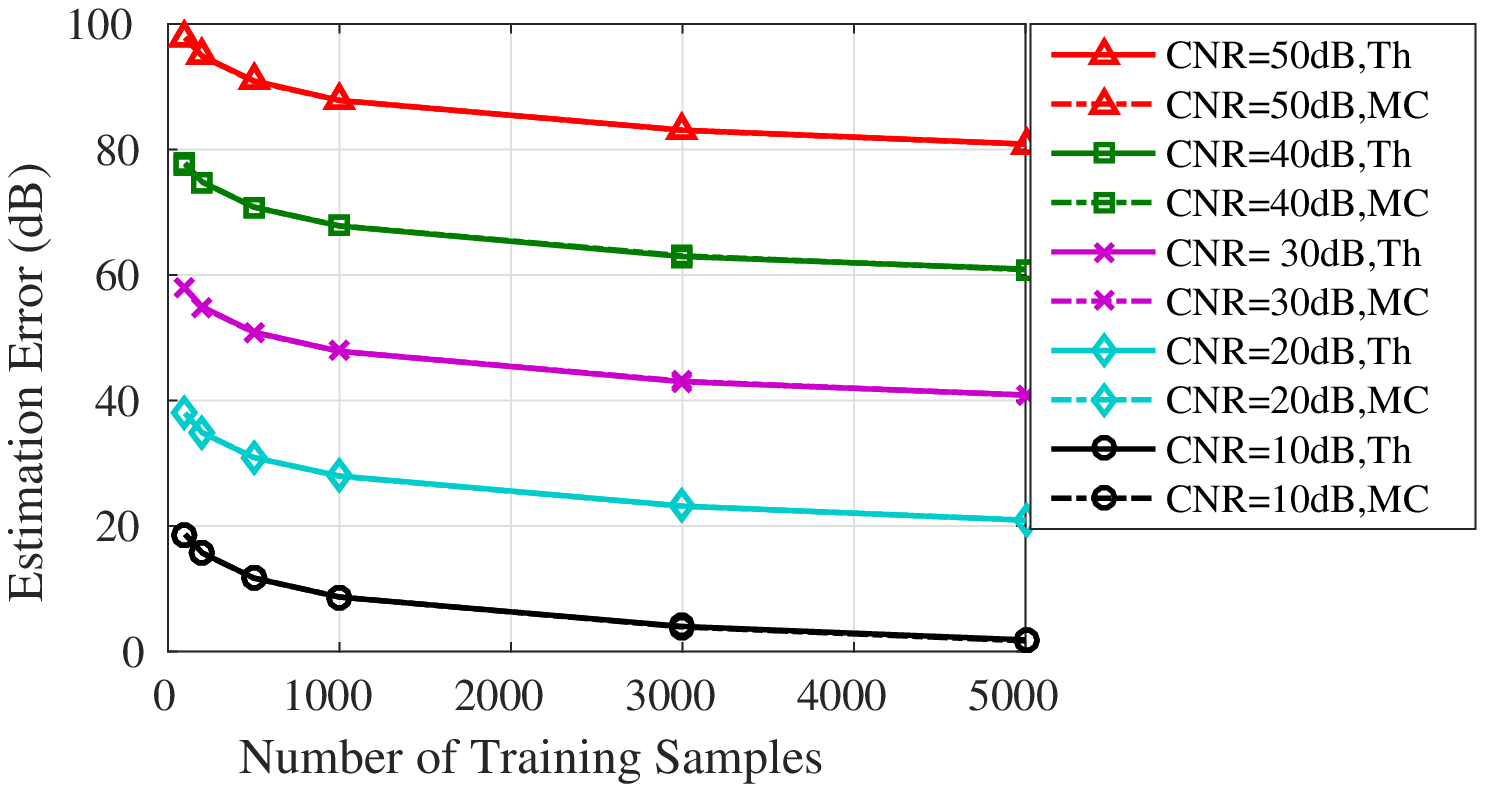}}
  \caption{\small Estimation errors of the virtual snapshots. (a) A small number of DoFs with $N_1=2$, $N_2=3$, and $M=8$. (b)  A median number of DoFs with $N_1=2$, $N_2=3$, and $M=20$. (c) A relatively large number of DoFs with $N_1=3$, $N_2=5$, and $M=20$. }\label{fig:ConvergenceVirSnp}
\end{figure}

In this example, we verify \eqref{biasDistrHat} using Monte Carlo experiments. Fig. \ref{fig:ConvergenceVirSnp} shows the variances of the estimated errors of virtual snapshots versus the number of samples at three selected system DoFs, i.e., (a) a small number of system DoFs with $N_1=2$, $N_2=3$, and $M=8$, (b) a median number of system DoFs with $N_1=2$, $N_2=3$, and $M=20$, (c) a relatively large number of system DoFs with $N_1=3$, $N_2=5$, and $M=20$. Five different $\textrm{CNR}$ values, i.e., $\textrm{CNR}=10\textrm{dB}$, $\textrm{CNR}=20\textrm{dB}$, $\textrm{CNR}=30\textrm{dB}$, $\textrm{CNR}=40\textrm{dB}$, and $\textrm{CNR}=50\textrm{dB}$ for each DoFs are considered. The solid curves indicate the theoretical values of the variances of the estimation errors given by $ \frac{1}{J} \sum^J_{i=1}\textrm{trace}({\bf \hat C}_i)$ ,where ${\bf \hat C}_i$ is calculated by \eqref{biasDistrHat} at $i$th Monte Carlo experiment, and the dashed curves show the results of Monte Carlo experiments given by $\frac{1}{J} \sum^J_{i=1} \| {\bf \hat z}_{r,i} - {\bf z}_r\|^2_2 $, where ${\bf \hat z}_{r,i}$ is the virtual snapshot estimate at $i$th Monte Carlo experiment and $J=1000$ is the total number of Monte Carlo experiments. Overall, it can be seen that, the theoretical values coincide with the results of Monte Carlo experiments. Moreover, it can be noted that the variance of the virtual snapshot estimation error significantly increases with the increase of the CNR values and DoFs, while it gradually decreases with the increase of the number of samples, and cannot reach zero even with a large number of samples or very small CNRs. This implies that the variance of the virtual snapshot estimation error is proportional to the CNR and DoFs, and inversely proportional to the number of samples, which can also be inferred from the theoretical derivation of \eqref{biasDistr1} and \eqref{biasDistrHat}.
\subsection{Performance of RTSKA-RD-SA-STAP} \label{Performance}
It is known that the accuracy of the prior knowledge would affect the performance of the knowledge-aided algorithms. In this subsection, numerical examples are provided to test the robustness against prior knowledge and consider the parameter setting issue of the RTSKA-RD-SA-STAP algorithm. It is assumed that the CPA has $N = 6$ sensors with coprime pair $N_1=2$ and $N_2 =3$. The number of pulses in one CPI is $M=18$, and $\textrm{CNR}=40$dB, and the number of training samples in each experiment is $5$. It is assumed that the measured values of platform velocity and crab angle $v'_p$ and ${\psi}'$ are uniformly distributed as within $v'_p \in [v_p - \Delta v_{pm}, v_p+\Delta_{pm} ] $ and $\psi' \in [\psi - \Delta \psi_{pm}, \psi + \Delta \psi_{m}]$, respectively, except for the fourth example. The $M_e$ is set to $15$ except for the second example. The range of azimuth angle is divided into $5N_v$ except for the third example. For comparison, the optimum performance of RTSKA-RD-SA-STAP with known covariance matrix (termed Proposed OPT) is shown.

\begin{figure}[!htbp]
\centering
  \includegraphics[width=85mm]{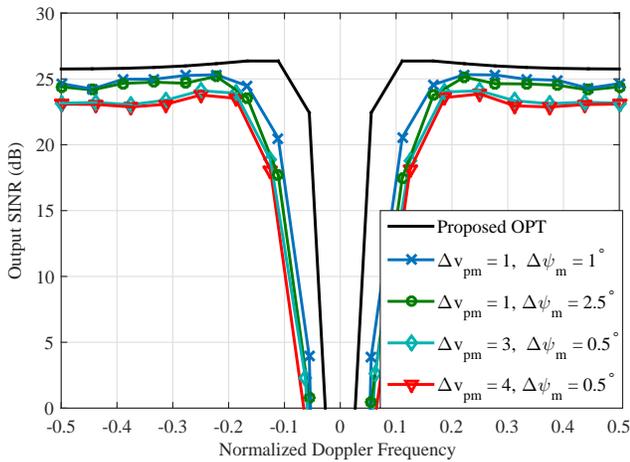}
  \caption{\small SINR comparisons for different prior knowledge errors with $N_1=2$, $N_2=3$, $M=18$, and $\textrm{CNR} = 40$dB. }\label{fig:Prior1}
\end{figure}

First, we evaluate the performance of the proposed RTSKA-RD-SA-STAP algorithm under different range values of errors in the crab angle and platform velocity. Specifically, we plot four different cases of prior knowledge errors: case 1, $\Delta v_{pm}=1 \textrm{m/s}$ and $\Delta \psi_{m} =1^\circ$; case 2, $\Delta v_{pm}=1 \textrm{m/s}$ and $\Delta \psi_{m} =5^\circ$; case 3, $\Delta v_{pm}=3 \textrm{m/s}$ and $\Delta \psi_{m} =0.5^\circ$; and case 4, $\Delta v_{pm}=4 \textrm{m/s}$ and $\Delta \psi_{m} =0.5^\circ$.
The resultant output SINRs versus the target normalized Doppler frequency are shown in Fig. \ref{fig:Prior1}. Obviously, as seen from the results, RTSKA-RD-SA-STAP is robust to the prior knowledge errors since the prior knowledge errors are considered in the design of RTSKA-RD-SA-STAP.

\begin{figure}[!htbp]
\centering
  \includegraphics[width=85mm]{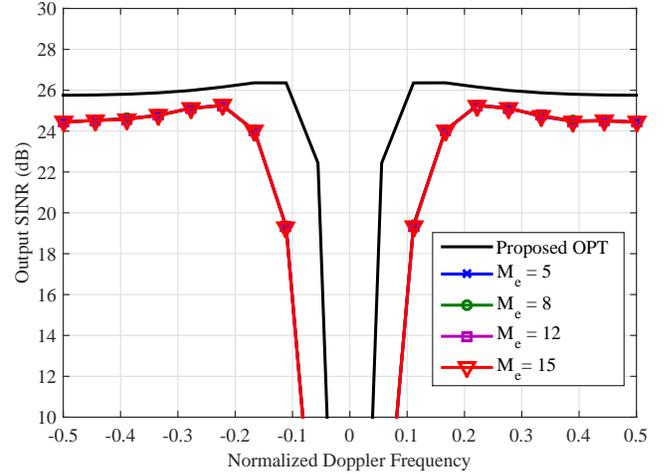}
  \caption{\small SINR comparisons for different values of parameter $M_e$ when $\Delta v_{pm} = 2 \textrm{m/s}$ and $\Delta \psi_{m} = 1^{\circ}$.}\label{fig:difNf}
\end{figure}

Next, the effect of value of $M_e$ on the performance of the proposed RTSKA-RD-SA-STAP algorithm is evaluated to provide a guideline for the setting of $M_e$. Here, $\Delta v_{pm} = 2$ and $\Delta {\psi}_{m} = 1^{\circ}$, and four different values of $M_e$, i.e, $M_e =5,8,12,15$, are considered as shown in Fig \ref{fig:difNf}. Obviously, it can be concluded that RTSKA-RD-SA-STAP is nonsensitive to the value of $M_e$. Therefore, we set $M_e$ to 15 in the following experiments for clarity.

\begin{figure}[!htbp]
\centering
  \includegraphics[width=85mm]{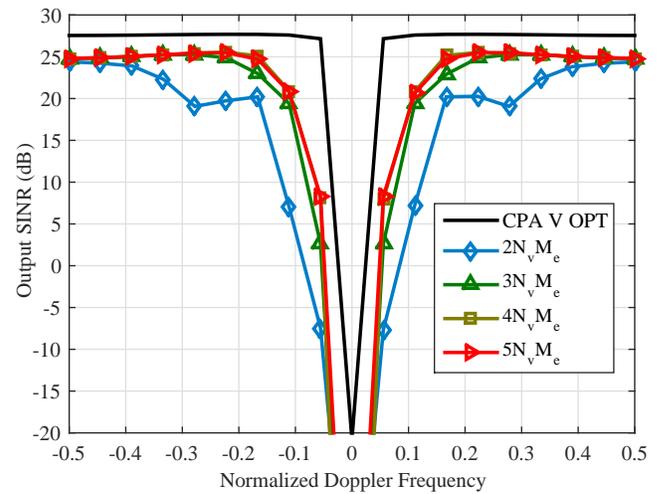}
  \caption{\small The impact of the number of dictionary atoms $M_d$ on the performance of the proposed RTSKA-RD-SA-STAP algorithm with known covariance matrix.}\label{fig:difMd}
\end{figure}

Third, the impact of the number of dictionary atoms (i.e. $M_d$) on the performance of RTSKA-RD-SA-STAP is evaluated. Four cases of number of dictionary atoms, i.e., $M_d = 2N_v M_e$, $M_d = 3N_v M_e$, $M_d = 4N_v M_e$, $M_d = 5N_v M_e$, are considered, and $\Delta v_{pm} = 2$ and $\Delta {\psi}_{m} = 1^{\circ}$. As seen from Fig. \ref{fig:difMd}, we observe that 1) the performance of RTSKA-RD-SA-STAP improves less when the number of dictionary atoms $M_d$ exceeds $4N_v M_e$, i.e. $M_d \geq 4N_v M_e$, 2) RTSKA-RD-SA-STAP fails to work for small number of the dictionary atoms, i.e. $M_d = 2N_v M_e$. Because there are serious mismatches between the overcomplete dictionary and the real clutter. These results are in accordance with the conclusions reported in \cite{Knowledge2006} and \cite{YangClut2013}. Therefore, the clutter subspace can be approximately accurately estimated by RTSKA-RD-SA-STAP when $M_d \geq 4N_v M_e$, and we set $M_d = 5N_vM_e $.

\begin{figure}[hbtp]
\centering
  \subfigure[SINR versus $\sigma_{\textrm{vp}}/\Delta \textrm{v}_{pm}$]{\label{fig:accuracy1}
  \includegraphics[width=83mm]{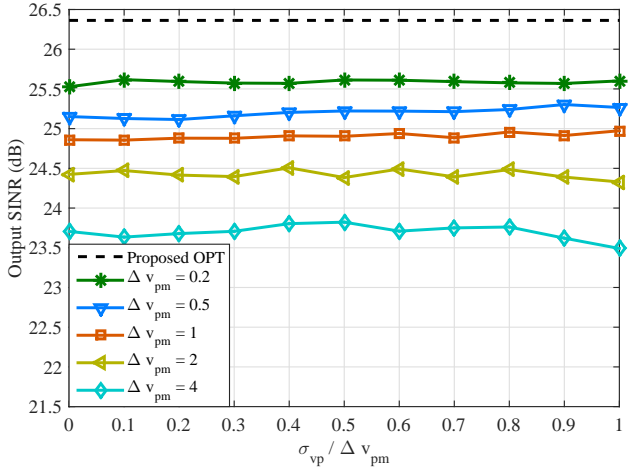}}
  \subfigure[SINR versus $\sigma_{\psi} / \Delta \psi_{m}$]{\label{fig:accuracy2}
  \includegraphics[width=83mm]{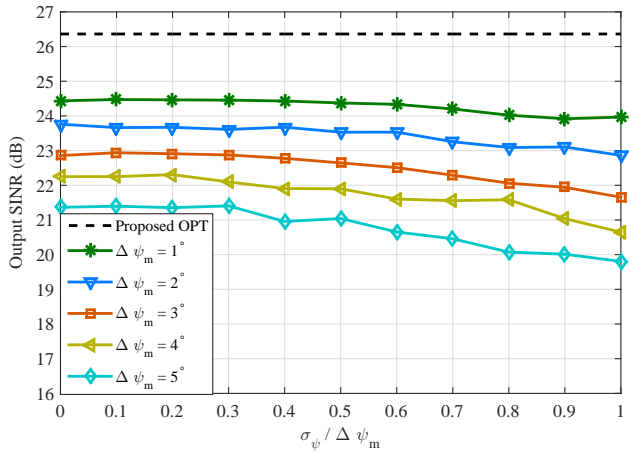}}
  \caption{\small SINR versus estimation accuracies of the platform velocity and crab angle. }\label{fig:accuracy}
\end{figure}

To further investigate the impact of the estimation accuracies of the range values of the platform velocity and the crab angle on the performance of RTSKA-RD-SA-STAP, in Fig. \ref{fig:accuracy1}, we show the output SINR of RTSKA-RD-SA-STAP versus the ratio of the standard variance and the mean of platform velocity error's range $\Delta {\textrm v}_{pm}$. Here, different from the above settings, both estimated error's range values of the platform velocity and the crab angle are modeled as Gaussian distribution with the mean of $\Delta {\textrm v}_{pm}$ and  $\Delta {\psi}_{m}$ and variance of ${\sigma}_{\textrm{vp}}$ and ${\sigma}_{\psi}$, respectively. The target normalized Doppler frequency is 0.1667. Obviously, it is observed that the performance degrades with the increase of the ratio ${\sigma}_{\textrm{vp}}/\Delta {\textrm v}_{pm}$, which is coincident with that reported by knowledged-aided algorithms \cite{Knowledge2006}. Moreover, RTSKA-RD-SA-STAP keeps constant output SINR when the ratio ${\sigma}_{\textrm{vp}}/\Delta {\textrm v}_{pm} \leq 0.6$, which is not a difficult task in practice. This implies that RTSKA-RD-SA-STAP is relatively robust to the estimation errors of the platform velocity. Moreover, in Fig. \ref{fig:accuracy2}, we plot the output SINR versus the the ratio of the standard variance and the mean of crab angle error's range $\Delta {\psi}_{m}$. From these results, it can be seen that the output SINR degrades gradually along with the increase of the ratio ${\sigma}_{\psi}/\Delta {\psi}_{m}$. On the other hand, when ${\sigma}_{\textrm{vp}}/\Delta {\textrm v}_{pm} =1$, RTSKA-RD-SA-STAP shows only 1.5dB SINR degradation. This suggests that RTSKA-RD-SA-STAP also offers robustness against the estimation errors of the crab angle. Therefore, RTSKA-RD-SA-STAP using the error's range values is more robust than the knowledge-aided algorithms directly using prior knowledge of the platform velocity and crab angle.

\begin{figure}[!htbp]
\centering
  \includegraphics[width=85mm]{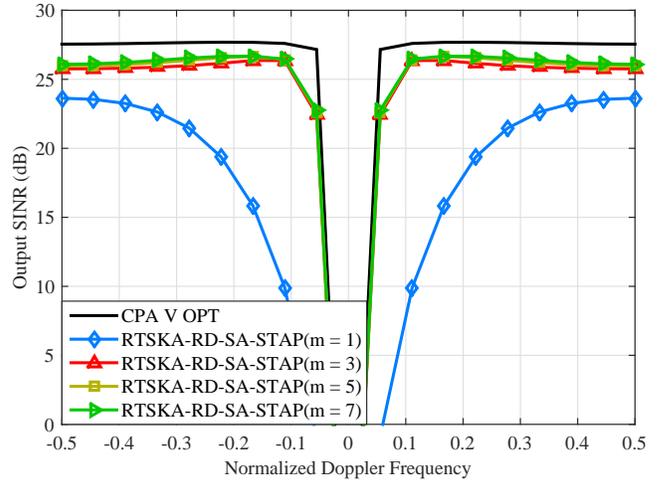}
  \caption{\small Proposed RTSKA-RD-SA-STAP algorithm performance for different number of RD channels $m$ with known covariance. } \label{fig:difchannel}
\end{figure}

Finally, the parameter setting issue on the number of Doppler channels selected is evaluated. Following the above settings, four cases of $m$ values, i.e. $m=1,3,5,7$, are considered, and $\Delta v_{pm} = 2$ and $\Delta {\psi}_{m} = 1^{\circ}$. Fig. \ref{fig:difchannel} shows the output SINRs versus the Doppler frequencies for different Doppler channels with known covariance matrix, where the virtual optimum performance corresponding to FD-SA-STAP with known covariance matrix (termed CPA V OPT) is also shown for comparison. It shows that the number of Doppler channels selected has significantly effect on the performance of RTSKA-RD-SA-STAP. It can be found that when $m=1$, RTSKA-RD-SA-STAP exhibits bad performance. However, when $m \geq3$, i.e., $m=3,5,7$, RTSKA-RD-SA-STAP can achieve better performance with about 2dB SINR loss within the optimal virtual counterpart and the performance for the $m=5$ and $m=7$ cases only slightly outperform that for the $m=3$ case. On the other hand, although the performance increases with the increase of the value of $m$, the computational complexity and sample support requirements also considerably increase with the increase of the value of $m$. Therefore, $m=3$ is a trade-off choice and is used in the following simulation examples.

\subsection{Comparison with Existing Algorithms}
\begin{figure}[!htbp]
\centering
  \includegraphics[width=85mm]{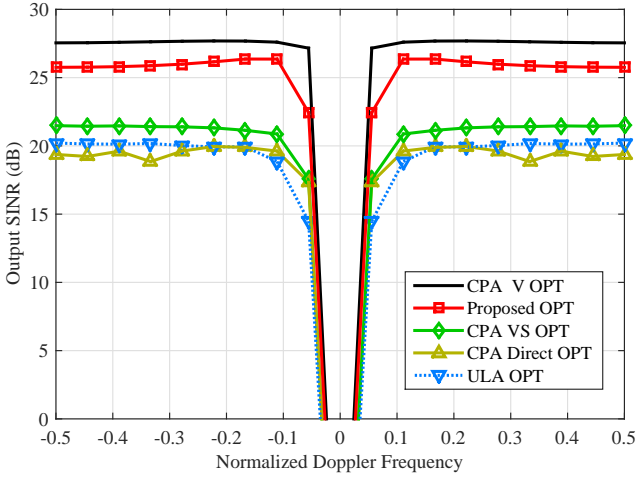}
  \caption{\small Theoretical performance comparisons and $m$=3, $N_1$=2, $N_2$=3, $M$=18, and $\textrm{CNR}=40$dB. }\label{fig:theoretical}
\end{figure}

We now compare RTSKA-RD-SA-STAP with existing algorithms including the FD-SA-STAP \cite{wxySparse2018}, VS-PC \cite{WangSTAP2018}, InAME-KA \cite{ZYEnhanced2017}, PC \cite{HaiEig1996}, JDL \cite{HWOnAdp1994}, mDT\cite{YLWVari2000}. At first, theoretical performance of mentioned above algorithms are compared when the covariance matrix ${\bf R}$ is known. The VS-PC algorithm with known covariance matrix (termed VS OPT) is shown. Moreover, the optimum performance corresponding to InAME-KA, PC, JDL, and mDT with known covariance matrix (termed Direct OPT) is also shown. The results are shown in Fig. \ref{fig:theoretical}. Here $N_1$ = 2, $N_2$ = 3, $M$ = 18,  $\textrm{CNR}$ = 40dB, $\Delta v_{pm} = 2$ and $\Delta {\psi}_m = 1^{\circ}$. It can be noticed that the proposed RTSKA-RD-SA-STAP algorithm exhibits the close performance to the virtual optimum performance (CPA V OPT), but significantly outperforms the other algorithms.

\begin{figure}[!htbp]
\centering
  \includegraphics[width=85mm]{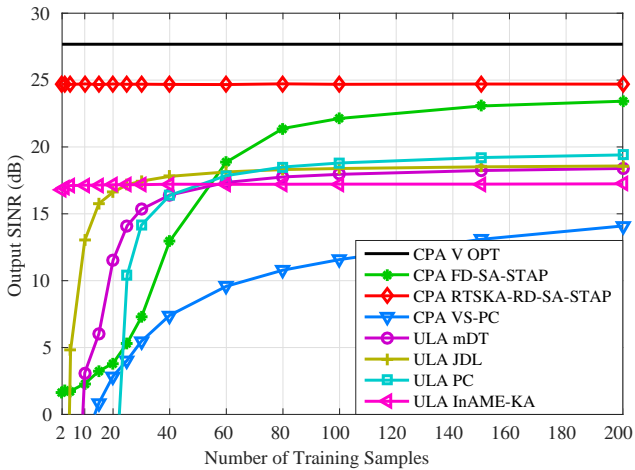}
  \caption{\small SINR versus the number of training samples. $N_1$=2, $N_2$=3, $M$=18, and $\textrm{CNR}=40$dB.}\label{fig:covergence}
\end{figure}

Next, we vary the number of training samples from 2 to 200, keep the target normalized Doppler frequency to 0.1667 and evaluate the convergence of RTSKA-RD-SA-STAP. Fig. \ref{fig:covergence} shows the SINR versus the number of training samples. It is seen that the FD-SA-STAP algorithm offers poor performance when the number of training samples is very small. However, the proposed RTSKA-RD-SA-STAP algorithm can perform well even when there is only one training samples and obtains the fastest convergence among all tested algorithms. Hence, the proposed RTSKA-RD-SA-STAP algorithm exhibits both good SINR performance and relatively low computational complexity for limited sample support. This is because the proposed RTSKA-RD-SA-STAP algorithm exploited the prior knowledge and the sparsity of the clutter.

\begin{figure}[!htbp]
\centering
  \includegraphics[width=85mm]{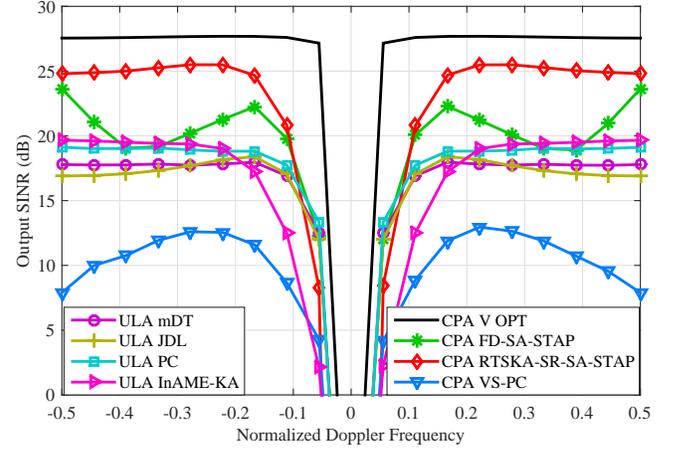}
  \caption{\small SINR for different target normalized Doppler frequencies. $N_1$=2, $N_2$=3, $M$=18, $L=100$, and $\textrm{CNR}=40$dB.}\label{fig:difDoppler}
\end{figure}

Moreover, we further compare the performance of the RTSKA-RD-SA-STAP algorithm with other algorithms of interest in terms of SINR performance versus different target Doppler frequencies, as shown in Fig. \ref{fig:difDoppler}. The number of training samples is set to $100$, and the other parameters are the same as those in the last example. Again, the FD-SA-STAP algorithm can not work well since the virtual FD virtual snapshot at high $\textrm{CNR}$ has large estimation errors. However, RTSKA-RD-SA-STAP is much more robust to the $\textrm{CNR}$ by exploiting the prior knowledge of the clutter.

\section{Conclusion} \label{Conclusion}
This paper has proposed a robust two-stage RD SA-STAP algorithm for airborne radar with CPAs considering inaccurate prior knowledge. The idea of RD preprocessing and inaccurate prior knowledge are introduced to tackle the performance degradation when the number of training samples is limited. Using the Toeplitz structure of the RD covariance matrix, an RD virtual snapshot was constructed and the relationship between the FD and the resultant RD virtual snapshot was established. An RD sparse signal model was also developed with the consideration of low complexity using the inaccurate prior knowledge of platform velocity and crab angle. The clutter subspace can be readily estimated by using an OMP-like method, where a robust method for estimating the clutter rank using these inaccurate knowledge was presented to guide parameter setting. Moreover, the convergence, practical implementations, and computational complexity of the proposed RTSKA-RD-SA-STAP algorithm were analyzed. Simulation results show a good estimate of clutter rank even for non-side looking array cases. It is also shown that RTSKA-RD-SA-STAP can offer robustness to a certain range of estimation errors of platform velocity and crab angle. Compared with the existing algorithms tested for the ULA with the same number of sensors as the CPA, RTSKA-RD-SA-STAP exhibits much better performance of clutter suppression. Compared with FD-SA-STAP algorithm, RTSKA-RD-SA-STAP can achieve even better performance in the case of a very small number of training samples with much lower complexity. In future research, we will consider the target-like interference, analyze the theoretical convergence, and assess the performance of RTSKA-RD-SA-STAP with real-world data.

\appendices
\section{Definition of the Matrix $\bf P$} \label{apdx1}
Similar to the coarray selection matrix definition reported in \cite{MianzhiWCMCRB2017}, the transformation matrix $\bf P$ creates the relationship between covariance vector and the virtual snapshot, which contains some holes rather than continuous samples. So $\bf P$ is an $N_v \times N^2$ matrix whose $k$th row and $l$th column entry satisfies
\begin{equation}\label{matrixP}
\begin{split}
{\bf P}_{k,l}=
\begin{cases}
\frac{1}{\omega (n_k)}& \text{if} \; d_{i}-d_{j}= n_k d_0,\\
0& \text{otherwise.}
\end{cases}
\end{split}
\end{equation}
where $k=1,2,\cdots, N_v$, $l= i+(j-1)N$ with $i=1,2,\cdots, N$ and $j=1,2,\cdots,N$, $\omega (n_k)$ denotes the number of all possible pairs $( d_i, d_j)$ such that $d_i-d_j = n_kd_0$, and $n_kd_0$ stands for the $k$th sensor position in the virtual array. A similar definition can be found in \cite{WangSTAP2018} and \cite{MianzhiWCMCRB2017}.

\section{ Definition of the Matrix $\bf F$ } \label{apdx2}
Using a similar idea as in \cite{WangSTAP2018}, the matrix $\bf F$ can be expressed as
\begin{eqnarray}\label{matrixF1}
\begin{split}
    {\bf F} = ( {\bf T} \otimes {\bf P} ) {\bf J},
\end{split}
\end{eqnarray}
where ${\bf T}$ and ${\bf P}$ are the transformation matrices for the slow time domain and space domain, respectively satisfying \cite{WangSTAP2018}
\begin{equation}\label{matrixF2}
\begin{split}
    {\bf b}_v(\varpi_{i,c}) = {\bf T} [ {\bf b}^*(\varpi_{i,c}) \otimes {\bf b}(\varpi_{i,c}) ] ,
\end{split}
\end{equation}
and
\begin{equation}\label{matrixF3}
\begin{split}
  {\bf a}_v(\vartheta_{i,c})  = {\bf P} [ {\bf a}^*(\vartheta_{i,c}) \otimes {\bf a}(\vartheta_{i,c}) ].
\end{split}
\end{equation}
Here, ${\bf P}$ is given by \eqref{matrixP}. Similar to the definition of $\bf P$ in \eqref{matrixP}, we first assume that the virtual pulse train at times $m_kT_r,k=1,\cdots,M_v$, then one gets
\begin{equation}\label{matrixPt}
\begin{split}
{\bf T}_{k,l}=
\begin{cases}
\frac{1}{\omega (m_k)}& \text{if} \; t_{i}-t_{j}=m_kT_r,\\
0& \text{otherwise.}
\end{cases}
\end{split}
\end{equation}
where $l= i+(j-1)M$ with $i=1,2,\cdots, M$ and $j=1,2,\cdots,M$, and $\omega (m_k)$ denotes the number of all possible pairs $( t_i, t_j)$ such that $t_i-t_j = m_kT_r$.

Moreover, ${\bf J}$ is an $N^2M^2 \times N^2M^2$ matrix and satisfies
\begin{equation}\label{matrixF4}
\begin{split}
   & [ {\bf b}^*(\varpi_{i,c}) \otimes {\bf b}(\varpi_{i,c}) ] \otimes [ {\bf a}^*(\vartheta_{i,c}) \otimes {\bf a}(\vartheta_{i,c}) ] \\
   & \quad \quad = {\bf J} [ {\bf b}^*(\varpi_{i,c}) \otimes {\bf a}^* (\vartheta_{i,c}) ] \otimes [ {\bf b}(\varpi_{i,c}) \otimes {\bf a} (\vartheta_{i,c}) ].
\end{split}
\end{equation}
And $\bf J$ is an invertible matrix with the $i$th row being all zeros except for a single 1 at the $j$th position, where $j$ is given by
\begin{equation}\label{matrixJ3}
\begin{split}
 j = (k4-1) N^2M + (k2-1)NM + (k3-1)N + k1.
\end{split}
\end{equation}
with
\begin{equation}\label{matrixJ2}
\begin{split}
& k4= \, \frac{i-1}{N^2M} + 1\!, \\
& k3 = \, \frac{\textrm{mod}(i-1,N^2M)}{N^2} \! +1, \\
& k2 = \, \frac{\textrm{mod}(\textrm{mod}(i-1,N^2M),N^2)}{N} \! +1, \\
& k1 =  \textrm{mod}(i-1,N)+1.
\end{split}
\end{equation}

%
%

\ifCLASSOPTIONcaptionsoff
  \newpage
\fi

\end{document}